\documentclass[twocolumn]{aastex701}

\usepackage{amsmath}
\usepackage{comment}

\usepackage{xcolor}

\definecolor{andrecolor}{RGB}{180,30,30} 

\begin{document}


\title{Elevated Eccentricities in the Radius Valley Hint at Water-Rich Mini-Neptunes}

\author[orcid=0000-0000-0000-0000,sname='Shibata']{Sho Shibata}
\affiliation{Department of Earth, Environmental and Planetary Sciences, MS 126, Rice University, Houston, TX 77005, USA}
\email[show]{s.shibata423@gmail.com}  

\author[orcid=0000-0003-1878-0634,sname='Izidoro']{Andre Izidoro}
\affiliation{Department of Earth, Environmental and Planetary Sciences, MS 126, Rice University, Houston, TX 77005, USA}
\email{ai18@rice.edu}


\begin{abstract}

While recent planet-formation models broadly reproduce the observed population of super-Earths and mini-Neptunes, as well as the bimodal radius distribution (the ``radius valley''), it remains unclear whether all these planets share a common rocky composition (a single popoulation of planets) or instead comprise two distinct populations --  rocky planets and icy planets (two populations of planets). The inferred eccentricity–radius relation, which shows a modest peak near the radius valley, provides a useful diagnostic for distinguishing between these scenarios. Here we use N-body simulations to examine how the radii and eccentricities of close-in planets depend on the masses and orbital configurations of their progenitor protoplanets.  We find that final planetary eccentricities scale with the system initial Safronov number. In two-population systems, energy equipartition between rocky and relatively more massive icy protoplanets creates a strong eccentricity contrast between the two groups, which appears as a peak near the radius valley. This signature does not appear if planetary systems are composed exclusively of rocky planets (with or without H-rich atmospheres), as assumed in photoevaporation and core-powered mass loss models. Because the eccentricity–radius relation traces a dichotomy in the underlying protoplanet mass distribution -- most plausibly arising from formation at different disk locations -- our results suggest that a significant fraction of mini-Neptunes are water-worlds. The observed radius and eccentricity distributions may reflect a mixture of systems that host exclusively rocky planets, systems dominated by icy planets, and systems with both rocky and icy planets.

\end{abstract}



\section{Introduction}

Among currently observed exoplanets, super-Earths and mini-Neptunes -- planets with radii of 1–4~$R_\oplus$ -- are the most common. Their size distribution is bimodal, exhibiting peaks near 1.4 and 2.4~$R_\oplus$, with a deficit at $\sim$1.8~$R_\oplus$ known as the ``exoplanet radius valley'' \citep{Fulton+2017, Fulton+2018}. The valley is widely interpreted as marking a compositional divide between two classes of planets \citep{Kuchner03, Fortney+2007, Adams+2008, Rogers+2010, Owen+2017}. 

In one view, both super-Earths and mini-Neptunes possess Earth-like rocky interiors, but mini-Neptunes retained thin H/He envelopes while super-Earths did not \citep{Owen+2017}. Alternatively, the radius valley is also consistent with predominantly rocky super-Earths and water-rich mini-Neptunes ({\it water worlds} or {\it ocean worlds}) \citep{Zeng+2019, Venturini+2020, Izidoro+2022, Burn+2024, Shibata+2025a, Nielsen+2025}. The true compositions of mini-Neptunes remain uncertain, rendering them a powerful testbed for planet-formation processes and key target for upcoming observational constraints \citep{Renyu+2025}.


 Recent transit light-curve analyses indicate that the eccentricity distribution of super-Earths and mini-Neptunes peaks near the radius valley at $\sim 1.8\,R_\oplus$ \citep{Gilbert+2025, Sagear+2025}. Planets within the valley exhibit median eccentricities of $\sim 0.1$ around FGK-type stars, whereas those outside the valley appear to display median eccentricities lower by a factor of $\sim$2--3. This suggests that planets in the valley have undergone stronger dynamical perturbations or orbital instabilities than those outside it. In this work we propose that the eccentricity–radius relation can be used to infer the compositional diversity of planets across the radius valley, revealing a connection that has not been previously recognized.

Analyses of orbital evolution in recent planet-formation models \citep{Shibata+2025b} show that a modest peak in eccentricity can arise when a sufficiently large number of rocky protoplanets (typically more than five) occupy the inner disk and the system also hosts outer icy mini-Neptunes capable of driving violent dynamical instabilities. These results suggest a potentially important link between inner rocky systems and outer water-rich planets -- one that may help explain both the exoplanet radius valley and its associated eccentricity structure. However, current simulations reproduce the observed eccentricity enhancement only under one particular model configuration. It remains unclear which aspects of planetary system architecture -- such as the number, masses,  and compositions -- are essential for generating this feature, and which configurations fail to do so. A clearer understanding of these dependencies requires a broader and more systematic exploration of parameter space.




To address this question, we perform a systematic suite of $N$-body simulations in which the number of planets, their masses, orbital configurations, and compositions are varied as free parameters. Rather than assuming a specific formation pathway, we remain agnostic about the origin of these architectures and instead aim to identify the physical conditions under which the observed eccentricity enhancement near the radius valley can be reproduced.


This paper is organized as follows. In Section~\ref{sec:methods}, we describe our methods. Sections~\ref{sec: single_population} and \ref{sec: two_population} present the results of our simulations for single- and two-population systems, respectively. In Section~\ref{sec: compare_obs}, we combine different simulation sets and compare our results with observational data. Section~\ref{sec: Safronov} discusses how the initial system configuration influences the final dynamical excitation. Finally, Section~\ref{sec: conclusion} summarizes our main conclusions.

\section{Methods}\label{sec:methods}
Our N-body simulations are performed using the \texttt{REBOUND} code \citep{Rein+2012} 
with the \textit{mercurius} integrator and an adaptive timestep set to \(1/40\) of the 
innermost protoplanet’s orbital period. We assume that, at the beginning of the simulations, 
the gaseous disk has already dissipated. We do not model gas-driven migration or the formation of resonant chains~\citep{Izidoro+2017,Izidoro+2022}. 

Formation models suggest that at the end of the gaseous diak phase, protoplanets are typically captured into mean-motion 
resonances, forming resonant chains \citep{Izidoro+2017}. The 
specific architecture of these chains -- particularly the sequence of resonances -- depends 
on the masses, number, and migration rates of protoplanets during their formation~\citep{Izidoro+2017,bitschizidoro24}. To reduce the number of free parameters in our study, we adopt a configuration in which protoplanets are initially spaced by equal orbital separations, corresponding to a fixed number of mutual Hill radii between adjacent bodies. This simplified setup allows us to better isolate and understand the physical processes of interest.

As mentioned above, our primary goal is to assess how the initial system configuration influences the emergence of the eccentricity peak reported by \citet{Gilbert+2025}. To this end, we conduct two types of simulations. The first corresponds to \textit{single-population systems}, in which stars hosts either only rocky planets or only icy planets. The second corresponds to \textit{two-population systems}, where rocky and icy planets coexist within the same system. In all our simulations, the central star is assumed to have one solar mass.

For single-population simulations we place $N_{0}$ protoplanets, each of mass $M_{0}$, in closely spaced orbits within the inner disk. Starting from an inner boundary at $a_{\mathrm{in,0}}$, the initial semi-major axes are assigned according to
\begin{align}
    a_{i+1,0} = a_{i,0} \frac{2+k_\mathrm{orb} h}{2-k_\mathrm{orb} h},
\end{align}
where
\begin{align}
    h = \left( \frac{M_{0,i+1}+M_{0,i}}{3 M_\odot} \right)^{1/3}.
\end{align}
where $a_{i,0}$ is the initial semi-major axis of the {\it i}-th protoplanet counted from the innermost orbit. We refer to the semi-major axis of the innermost protoplanet as $a_\mathrm{in,0}:=a_{0,\mathrm{0}}$. We set the orbital separation factor to \(k_{\mathrm{orb}} = 10\) in our main simulations. In our simulations, the composition of protoplanets are set to rocky (Earth-like composition) or icy (50\% water and 50\% Earth-like). Their radii are obtained from the mass--radius relation of \citet{Zeng+2019} for these specific compositions. The initial orbital eccentricities and inclinations of protoplanets are randomly sampled from a Rayleigh distribution, assuming energy equipartition of $\langle e^2 \rangle^{1/2} = 2 \langle \sin^2 i \rangle^{1/2} = h$. The remaining orbital angles are randomly selected from a uniform distribution between \(0^\circ\) and \(360^\circ\).

In our two-population simulations, in addition to $N_0$ rocky protoplanets of mass $M_0$, we include $N_{\mathrm{ext},0}$ external protoplanets, each with mass $M_{\mathrm{ext},0}$. These external bodies represent icy protoplanets, and their semi-major axes, eccentricities, inclinations, and other orbital elements are assigned using the same procedure adopted for protoplanets in single-population simulations.

We also investigated how the choice of initial orbital spacing affects the subsequent orbital evolution of protoplanets (Appendix~\ref{sec: orbital_separation}). We note for the reader that the resulting eccentricity distribution depends only weakly on the protoplanets’ initial mutual separations.

We integrate each system for up to \(2 \times 10^{9} P_{\mathrm{in,0}}\), where 
\(P_{\mathrm{in,0}}\) is the initial orbital period of the innermost protoplanet. 
Orbital-crossing timescales for equally spaced planetary systems suggest that orbital instabilities typically begin within the characteristic ages of planetary systems in our Galaxy (\(\lesssim 1\)~Gyr; \citealt{Zhou+2007}). After each integration segment is completed, we check whether the system has undergone its orbital instability. If no giant impacts have occurred, or if the last giant impact occurred within the most recent \(1 \times 10^{9} P_{\mathrm{in,0}}\) segment, we consider the system potentially still unstable and extend the integration by an additional \(1 \times 10^{9} P_{\mathrm{in,0}}\). We repeat this process, extending the integration in sequential steps as needed, until each system has fully completed its orbital instability.




Following the completion of our N-body simulations, we perform a post-formation analysis to determine planetary radii. We assume that each planet have a primordial (H-rich) atmosphere and compute the final radius $R$, assuming an initial atmospheric mass-fraction of $0.3\%$ of the core mass. 
We then model atmospheric mass loss due to photoevaporation, as well as the resulting evolution of $R$ from the end of the $N$-body simulations to 3~Gyr, using the prescription of \citet{Owen+2017}. An illustrative example of the photoevaporation evolution is shown in Appendix~\ref{sec: Photoevapolation}. Note that in our $N$-body simulations we neglect the explicit presence of planetary atmospheres during the dynamical evolution and collisions. During the $N$-body simulations all protoplanets are treated as objects whose masses include a small contribution from their primordial gaseous envelopes  -- assumed to correspond to 0.3\% of their individual mass.  Collisions are modeled as perfect mergers that conserve the total mass, and we do not track atmospheric stripping during impacts; in other words, the envelope mass is assumed to remain bound to the merged body and is only modified later in the post-formation analysis. Because the envelope masses are much smaller than the core masses  they do not affect the orbital evolution in any significant way.

In addition to photoevaporation, atmospheric mass loss can also be triggered by giant impacts through shock propagation \citep{Roche+2025}, and can be further enhanced by core-powered mass loss \citep{Biersteker+2019, Biersteker+2020, Darius+2025}. Post-impact mass-loss rate depends on the efficiency of energy deposition into the mantle, the core’s cooling rate, and the subsequent thermal evolution of the atmosphere. As mentioned before, for simplicity, we neglect impact-induced mass loss in our main analysis. Importantly, in Appendix~\ref{app: impact_induced_mass_loss} we show that including this effect does not alter our main conclusions.

\section{Single-population systems}\label{sec: single_population}


We treat the initial number of protoplanets $N_\mathrm{0}$, their initial mass $M_\mathrm{0}$, the initial location of the innermost protoplanets $a_{\mathrm{in},0}$, and the planetary composition as free parameters. For each parameter set, we perform 50 simulations using different random seeds. The full list of parameter sets considered in this section is summarized in Table~\ref{tab: parameters_for_single}.

\begin{table}[h]
    \centering
    \begin{tabular}{|c|c|c|c|c|}
    \hline
    Sim. ID & $a_\mathrm{in,0}$ & $N_\mathrm{0}$ & $M_\mathrm{0}$  & Composition \\ 
    \hline
    \hline
    S00  & $0.05$ au   & 10    & $[0.5-4.0] M_\oplus$ & rocky \\  
    S01  & $0.10$ au   & 10    & $[0.5-4.0] M_\oplus$ & rocky \\  
    S02  & $0.05$ au   & [6-22]   & $0.5 M_\oplus$ & rocky \\  
    S03  & $0.10$ au   & [6-22]   & $0.5 M_\oplus$ & rocky \\  
    S04  & $0.05$ au   & [6-22]   & $1.0 M_\oplus$ & rocky \\  
    S05  & $0.10$ au   & [6-22]   & $1.0 M_\oplus$ & rocky \\  
    S06  & $0.05$ au   & [6-14]   & $4.0 M_\oplus$ & icy \\  
    S07  & $0.10$ au   & [6-14]   & $4.0 M_\oplus$ & icy \\  
    S08  & $0.05$ au   & [6-14]   & $8.0 M_\oplus$ & icy \\  
    S09  & $0.10$ au   & [6-14]   & $8.0 M_\oplus$ & icy \\  
    \hline
    \end{tabular}
    \caption{Parameters defining the configurations of single-population protoplanetary systems. Planets are either assumed rocky or icy.}
    \label{tab: parameters_for_single}
\end{table}

\subsection{Orbital evolution of a single-population system}

\begin{figure}
    \centering
    \includegraphics[width=0.9\linewidth]{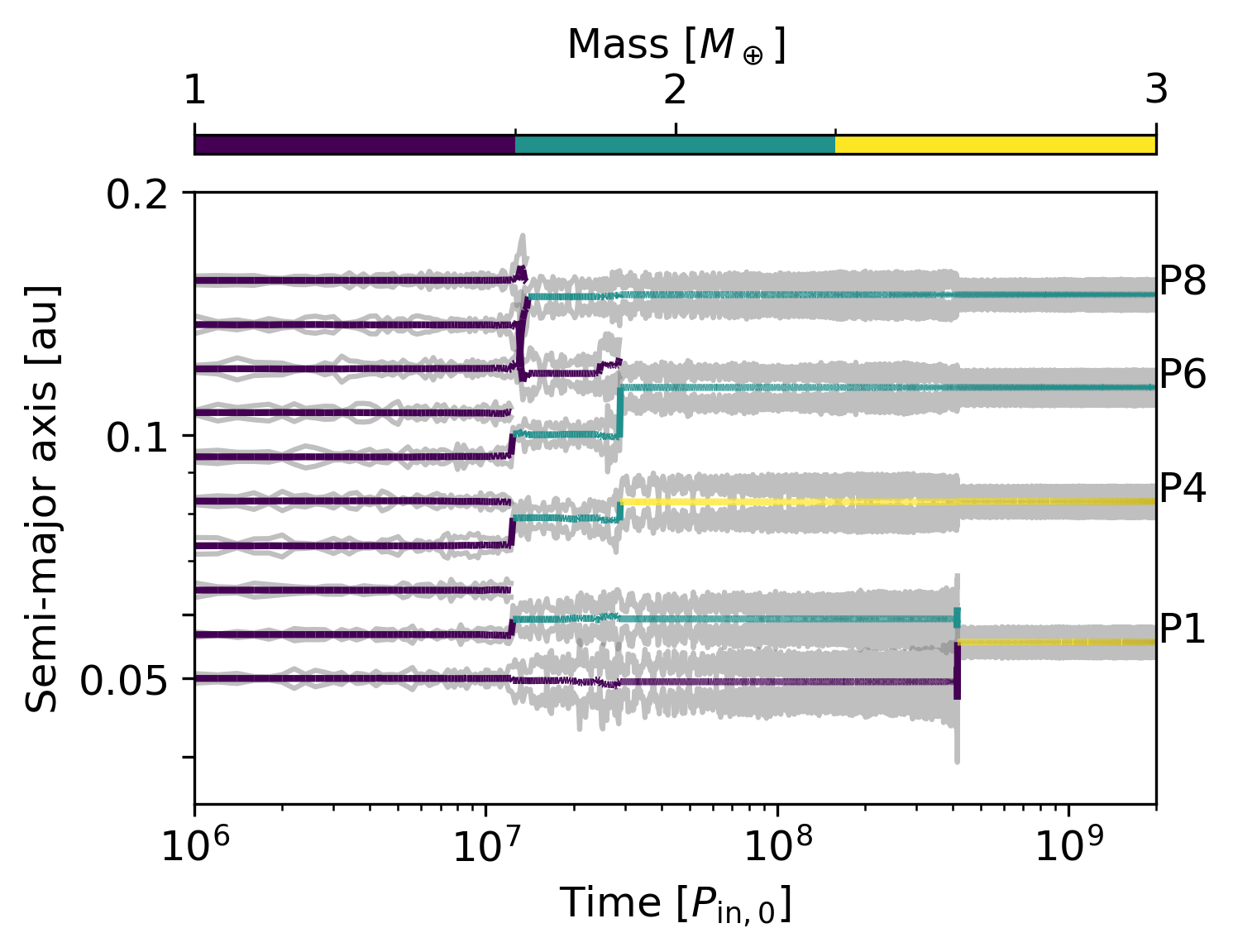}
    \caption{
    Time evolution of the protoplanet's semi-major axis in one representative simulation with $N_{0}=10$, $M_{0}=1 M_\oplus$, and $a_{0,0}=0.05,\mathrm{au}$. The color-coding shows the planetary mass. Gray lines show the pericenter and apocenter of protoplanets' orbits. The IDs of each protoplanet are shown on the right side of the panel.}
    \label{fig: orb_single}
\end{figure}

Figure~\ref{fig: orb_single} shows a representative example of the dynamical evolution in one of our simulations, in which all planets are initially assumed to be rocky. In this case, 10 protoplanets with initial mass $M_{\mathrm{0}} = 1.0\,M_\oplus$ are placed with the innermost body at $a_{\mathrm{in},0} = 0.05$\,au. The eccentricities gradually increase, and the system enters the orbital instability phase around $t = 10^{7} P_{\mathrm{in},0}$. Even after the first giant impacts, eccentricities continue to grow and further collisions occur. Following the collisions at $4\times10^{8}P_{\mathrm{in},0}$, the orbits become well separated and the system transitions to a long-term stable phase.

\subsection{The effect of the initial mass of protoplanets $M_0$}\label{sec: e-Rp_single}

\begin{figure}
    \centering
    \includegraphics[width=1.0\linewidth]{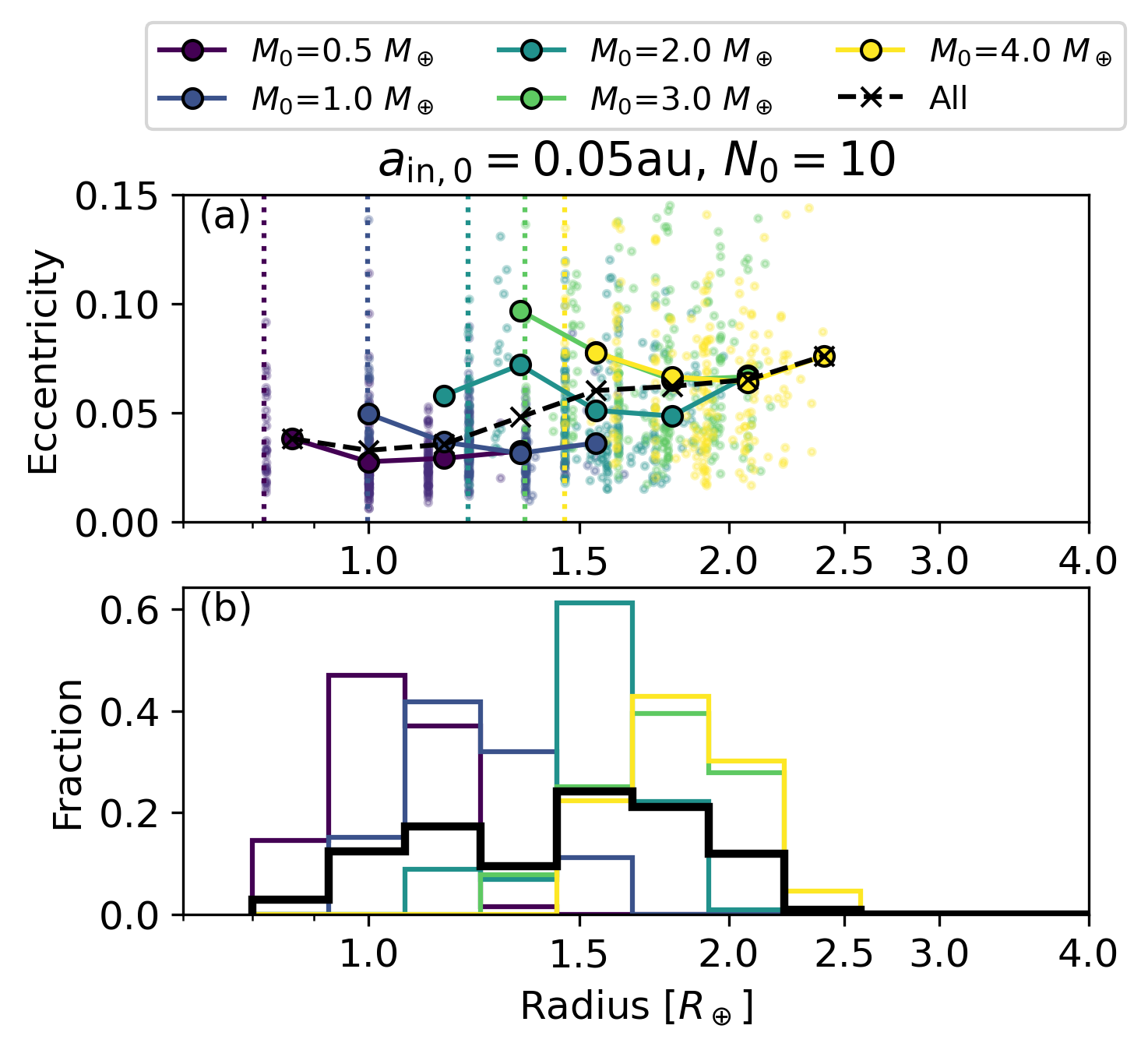}
    \caption{
    {\bf Panel-(a)}: mean eccentricity of planets as a function of their physical radius. Filled circles connected with solid lines show the mean eccentricity of planets in each bin. Different colors correspond to scenarios with different initial masses of protoplanets $M_\mathrm{0}$, as shown in the legend box. The black dashed line combines all simulations.     {\bf Panel-(b)}: histogram showing the distribution of planetary radius. The bin size is the same as that in panel (a).  Scenarios corresponds to cases where $N_\mathrm{0}=10$ and $a_\mathrm{in,0}=0.05$ au.
    }
    \label{fig: Rp_Ecc_single_psMp}
\end{figure}

We start by discussing the results obtained from simulations with different initial protoplanet masses $M_\mathrm{0}$, while keeping $N_\mathrm{0}=10$ and $a_{\mathrm{in},0}=0.05$\,au fixed (the simulations labeled S00 in Table~\ref{tab: parameters_for_single}). Figure~\ref{fig: Rp_Ecc_single_psMp}(a) shows the distributions of planetary radii and eccentricities. In the horizontal axis, we consider  13 logarithmically spaced bins between $R=0.7\,R_\oplus$ and $4.0\,R_\oplus$, and restrict the plotted sample to short-period planets with $P_\mathrm{orb}<100$\,days. Because secular interactions continue to drive eccentricity oscillations even after the planets' orbits are well separated, we characterize the final eccentricity by averaging over the last 5~Myr of the simulations. Throughout this paper, the quantity $e$ therefore refers to this time-averaged eccentricity. For each radius bin, we compute the mean (time-averaged) eccentricity $\langle e \rangle$ and plot the corresponding $\langle e \rangle$--$R$ relation using solid lines.
Figure~\ref{fig: Rp_Ecc_single_psMp}(b) shows the distribution of planetary radii presented as a histogram.

\begin{figure}[h]
    \centering
    \includegraphics[width=0.9\linewidth]{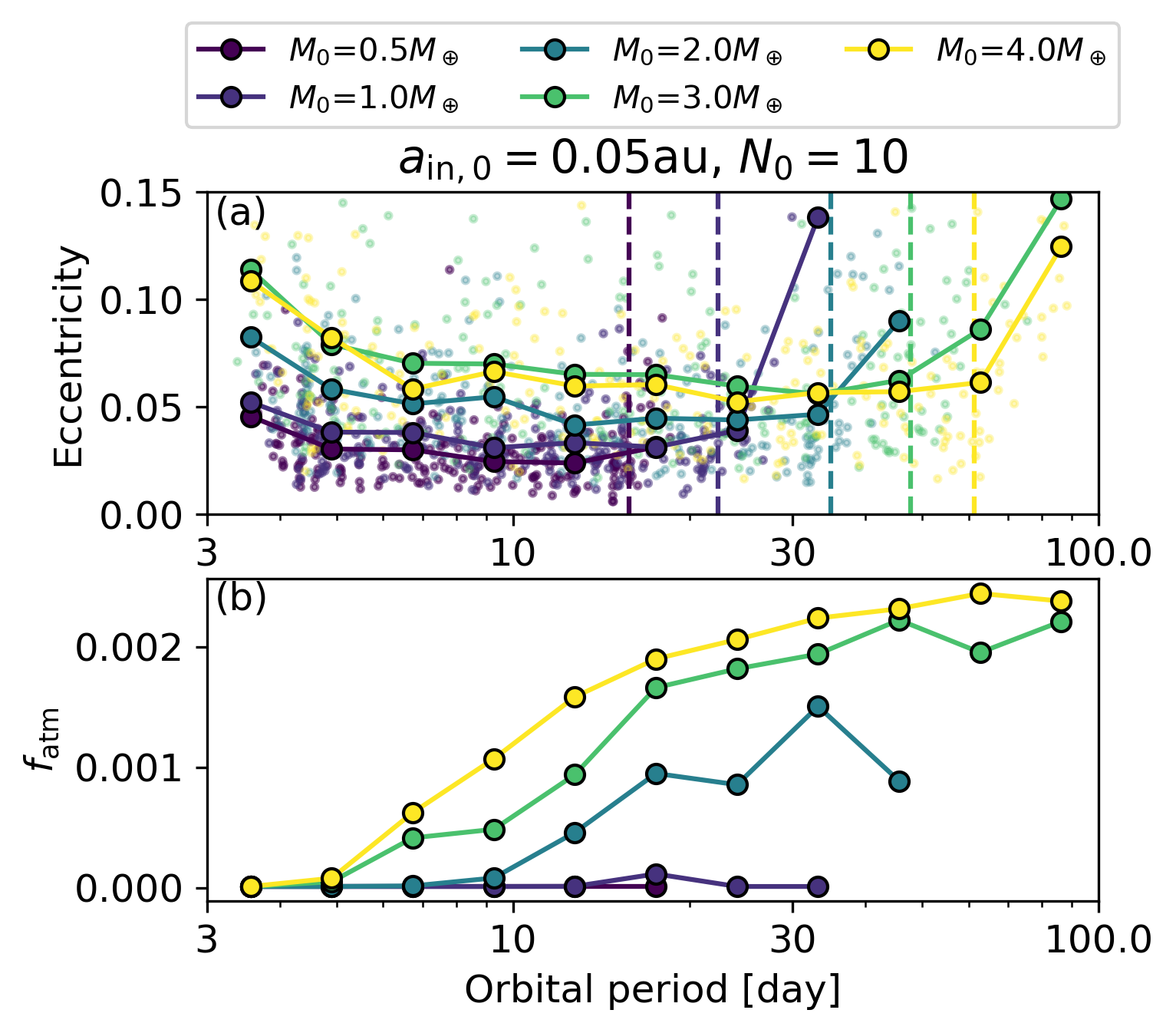}
    \caption{
    {\bf Panel-(a)}: mean eccentricity of planets as a function of their orbital period. Filled circles connected with solid lines show the mean eccentricity of planets in each bin. We use 11 logarithm  bins between 3--100 days. Different colors correspond to cases with different initial masses of protoplanets $M_\mathrm{0}$, as shown in the legend box. Vertical dashed lines show the semi-major axis of the outermost protoplanet at the beggining of the simulations. Scenarios corresponds to cases where $N_\mathrm{0}=10$ and $a_\mathrm{in,0}=0.05$ au.
    {\bf Panel-(b)}: distribution of the final averaged atmosperic mass fraction $f_\mathrm{atm}$. The bin size is the same as in panel (a).
    }
    \label{fig: Per_Ecc_single_psMp}
\end{figure}


First, we focus on the shape of $\langle e \rangle$--$R$ curves obtained in each simulation. Although each $\langle e\rangle$–$R$ curve is nearly flat in Fig.~\ref{fig: Rp_Ecc_single_psMp}(a), there is a slight upward trend toward small radii. To understand the reason behind, we plot the eccentricities against their orbital period in Fig.~\ref{fig: Per_Ecc_single_psMp}(a). The obtained $\langle e \rangle$--$P$ curves have bowl-shaped profiles. For protoplanets with the intermediate orbital-period (e.g., around ten days), strong scattering typically leads to collisions with adjacent protoplanets. Because giant impacts themselves act as an efficient eccentricity-damping process \citep{Matsumoto+2015}, a sequence of scattering events followed by collisions tends to maintain the planet’s eccentricity at a moderate level. However, the scattered innermost and outermost protoplanets tend to avoid collisions and sustain higher eccentricities, which results in the bowl-shaped forms in $\langle e \rangle$--$P$ relations. Since those planets undergo fewer giant impacts, the smallest planets in the system tend to retain slightly higher eccentricities than the others.

Next, we focus how the initial planetary mass affects the eccentricity distributions. Figure~\ref{fig: Rp_Ecc_single_psMp}(a) shows that the mean final eccentricity increases with the initial planetary mass $M_0$. This is also seen in Figure~\ref{fig: Per_Ecc_single_psMp}(a). This behavior is expected, because planet--planet scattering becomes more vigorous in systems containing more massive protoplanets. Figure~\ref{fig: Rp_Ecc_single_psMp}(b) shows the corresponding radius distribution. When $M_0 \lesssim 1\,M_\oplus$, almost all planets lose their atmospheres due to photoevaporation (see also Fig.~\ref{fig: Per_Ecc_single_psMp}(b) showing the mean atmospheric mass fraction $f_\mathrm{atm}$), and the radius distribution peaks at $1$--$1.2\,R_\oplus$. In contrast, when $M_0 \gtrsim 3\,M_\oplus$, most planets (except those on very close-in orbits) retain their atmospheres (see Fig.~\ref{fig: Per_Ecc_single_psMp}(b)), resulting in planets with radii larger than $1.5\ R_\oplus$. In summary, the peak of the radius distribution shifts to larger radii with increasing $M_0$ because both the core masses and the atmospheric mass fractions increase.

\citet{Owen+2017} showed that a population of rocky planets with characteristic masses of $\sim3 M_\oplus$ can broadly reproduce the observed radius valley. When we combine all simulations in S00, we likewise obtain a radius distribution that exhibits a modest gap (black solid line in Fig.~\ref{fig: Rp_Ecc_single_psMp}(b)). The location of this shallow gap differs slightly from that found by \citet{Owen+2017} and from observations, because our simulations produce a somewhat different distribution of core masses. Matching the exact position of the radius valley is not our main objective here.

More importantly,  when combining all S00 simulations, the mean eccentricity increases smoothly across the gap location and shows no indication of a peak in eccentricity (black dashed line in Fig.~\ref{fig: Rp_Ecc_single_psMp}(a)). This result is robust and does not depend sensitively on the precise location of the radius valley. We therefore conclude that systems composed of rocky planets only do not reproduce the observed eccentricity enhancement near the radius valley.


\subsection{The effect of the initial number of protoplanets $N_0$}

\begin{figure}
    \centering
    \includegraphics[width=0.9\linewidth]{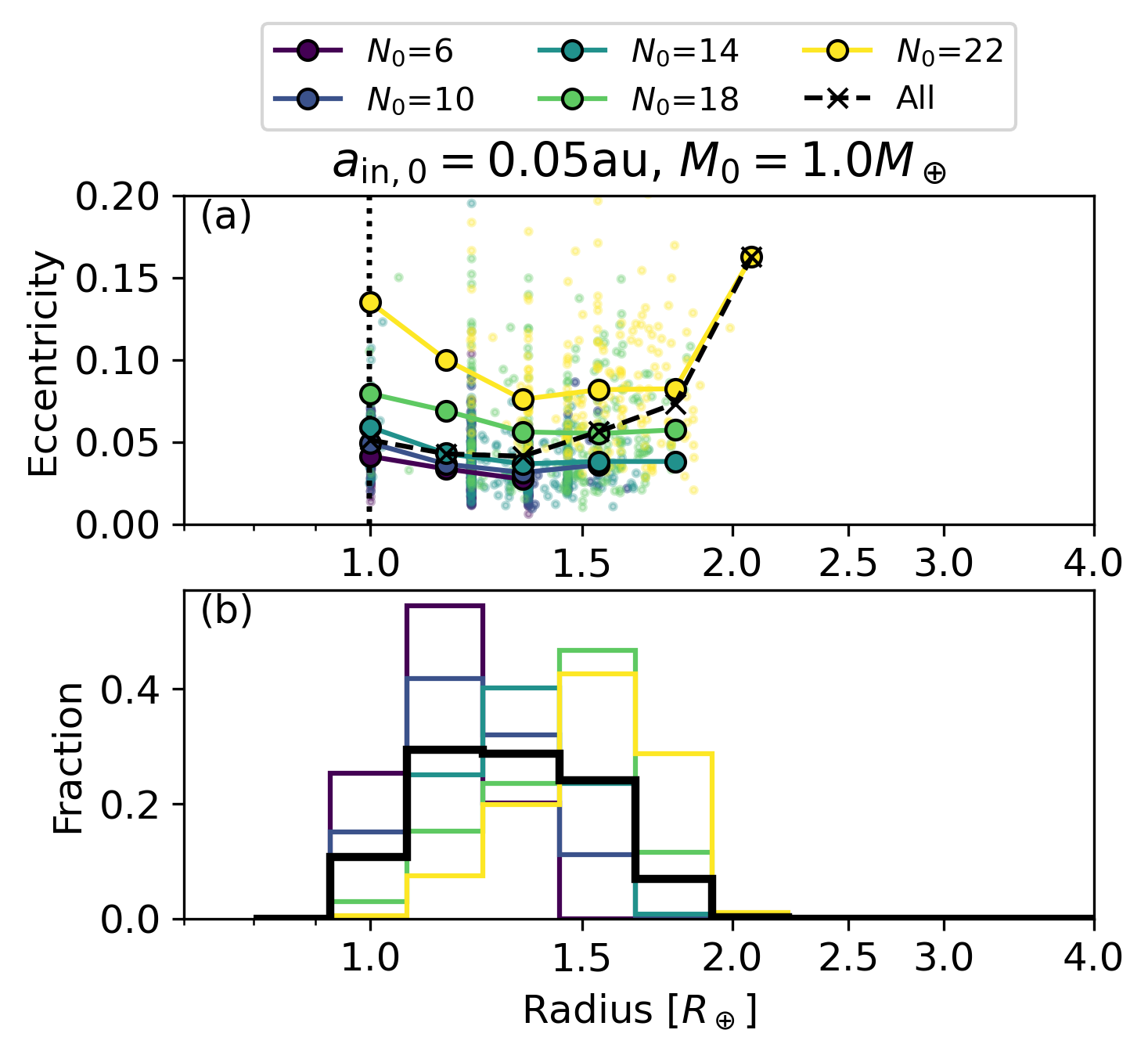}
    \caption{
    Same as Fig.~\ref{fig: Rp_Ecc_single_psMp}, but for scenarios considering a different initial number of protoplanets. It corresponds to  scenarios cases where $M_\mathrm{0}=1M_\oplus$ and $a_\mathrm{in,0}=0.05 \mathrm{au}$. The vertical dotted line shows the  core radius at the beginning of the simulations.}
    \label{fig: Rp_Ecc_single_psNp}
\end{figure}

\begin{figure}
    \centering
    \includegraphics[width=0.9\linewidth]{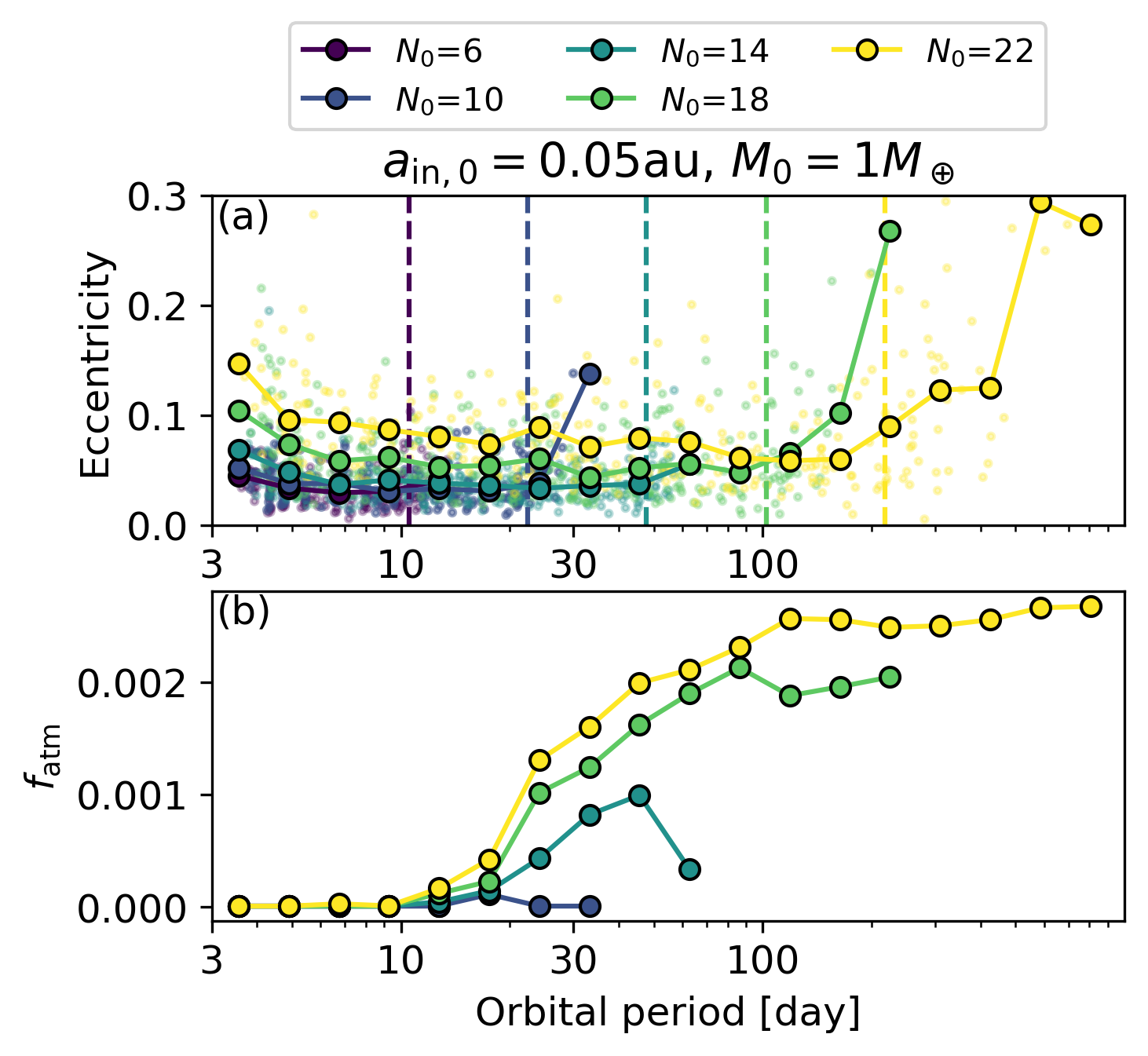}
    \caption{
    Same as Fig.~\ref{fig: Per_Ecc_single_psMp}, but for a set of scenarios considering different initial protoplanet number $N_0$. The bin size is same as Fig.~\ref{fig: Per_Ecc_single_psMp}, but extended to $P=1000$ days.
    }
    \label{fig: Per_Ecc_single_psNp}
\end{figure}

We next analyze our simulations carried out with different initial numbers of protoplanets $N_\mathrm{0}$, while fixing $M_\mathrm{0}=1 M_\oplus$ and $a_\mathrm{in,0}=0.05$ au (the simulations labeled S04 in Table~\ref{tab: parameters_for_single}). This setup is motivated by formation models suggesting that the number of protoplanets in the inner disk depends on the available solid mass, migration timescales, and the disk lifetime \citep{Shibata+2025b}.

Similar to the S00 results (Fig.~\ref{fig: Rp_Ecc_single_psMp}), the $\langle e \rangle$–$R$ curves in Fig.~\ref{fig: Rp_Ecc_single_psNp}(a) show a slight decrease in mean eccentricity with increasing planetary radius, although some bins are affected by small-number statistics (see bins at smallest and largest radii for $N_0=22$. These bins contain less than 1 \% of the all planets in these simulations.). Another clear trend is that the overall eccentricity level tend to increase as the initial number of protoplanets increases.

Figure~\ref{fig: Rp_Ecc_single_psNp}(b) shows that the planetary radius distribution shifts toward larger radii as $N_0$ increases. The corresponding radius distribution is single-peaked and does not exhibit a radius valley. From Fig.~\ref{fig: Rp_Ecc_single_psNp}, we find that both the mean eccentricities and the mean radii of planetary systems increase with $N_0$. As before, when combining all simulations (black dashed line), we obtain an overall $\langle e \rangle$--$R$ trend in which the mean eccentricity increases toward larger radii.

Figure~\ref{fig: Per_Ecc_single_psNp}(a) shows the resulting $\langle e \rangle$–$P$ curves. Notably, all simulations begin with the same number of protoplanets interior to $P<10$ days. However, systems with larger $N_0$ generate higher eccentricities overall because they host ``additional'' protoplanets further out. A clear example is the comparison between the $N_0=18$ and $N_0=22$ cases: the only difference is the presence of four extra protoplanets at $P\sim100$~days, yet these additional bodies noticeably enhance the eccentricities of planets with $P<10$~days. This is interesting because it indicates that the level of eccentricity excitation is not governed solely by interactions among nearby neighbors; gravitational perturbations from more distant planets also play a significant role. Note that increasing $N_0$ also increases the system’s total mass. However, as we show later in the paper, for systems with the same total initial mass, the final level of eccentricities correlates with $N_0$ (see Sec.~\ref{sec: Safronov}).

\subsection{The effect of planetary composition}\label{sec: single_composition}

\begin{figure}
    \centering
    \includegraphics[width=1.0\linewidth]{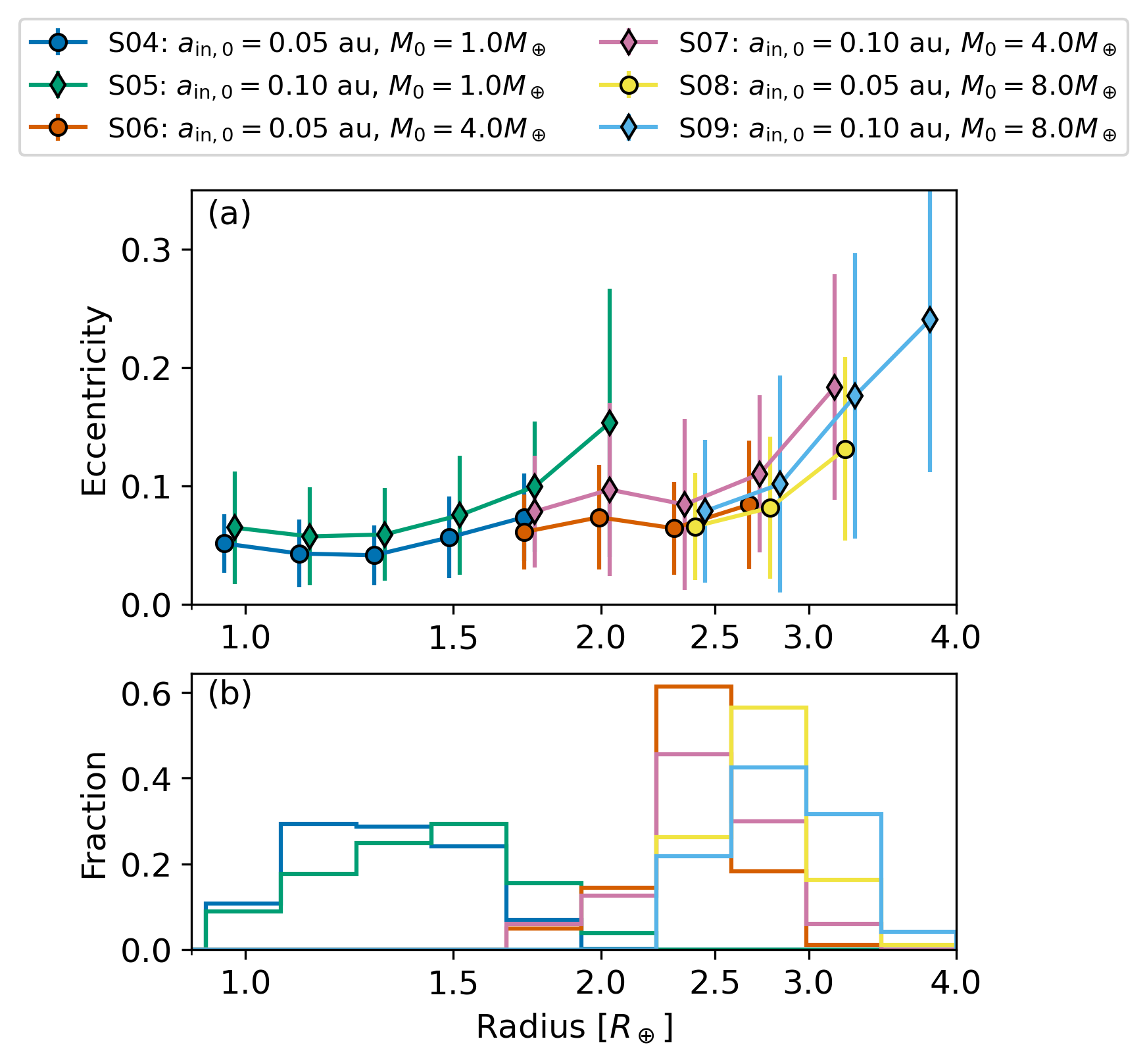}
    \caption{
    Same as Fig.~\ref{fig: Rp_Ecc_single_psNp}, but for a different set of simulations. We combine the simulations in each parameter set, and each solid line corresponds to a simulation ID, as labeled in the legend box. The error bars show the standard deviations. The circle and diamond plots show the simulations with $a_\mathrm{in,0}=0.05$ au and $0.10$ au, respectively. To avoid biases from small-number statistics, we remove plots whose subsamples comprise less than 2\% of the full dataset.
    }
    \label{fig: Rp_Ecc_single_combined}
\end{figure}



Until now, we have focused on systems composed exclusively of rocky planets (with primordial atmospheres that may be lost due to photoevaporation). We next turn to scenarios involving planets of different compositions (Fig.~\ref{fig: Rp_Ecc_single_combined}), in order to systematically compare them.

Because icy protoplanets form beyond the water ice line and must be sufficiently massive to migrate inward, we initialize them with larger masses than rocky protoplanets ($M_0 = 4M_\oplus$ or $8M_\oplus$). This choice is also consistent with the fact that the pebble-isolation mass increases with orbital distance \citep{Lambrechts+2014b, Bitsch+2018}. Moreover, formation models show that planets formed beyond the water ice line are typically more massive than those formed interior to it regardless of how planets grow, as for instance via pebble or planetesimal accretion~\citep[e.g.,][]{levisonetal15, izidoroetal21,Shibata+2025a}.

Scenarios S04–S05 contain exclusively rocky protoplanets, whereas scenarios S06–S09 contain exclusively icy protoplanets. We compare the $\langle e \rangle$–$R$ distributions obtained in these specific scenarios of Tab.~\ref{tab: parameters_for_single} in Fig.~\ref{fig: Rp_Ecc_single_combined}. Circles and diamonds represent simulations with $a_{\mathrm{in},0}=0.05$ au and $a_{\mathrm{in},0}=0.10$ au, respectively. Note that for each of these scenarios we performed simulations considering different $N_0$. For a given scenario (Sim. ID in Tab.~\ref{tab: parameters_for_single}), we combine the results of all simulations with different $N_0$. In other words, the blue line in Fig.~\ref{fig: Rp_Ecc_single_combined}(a) is the same as the black dashed line in Fig.~\ref{fig: Rp_Ecc_single_psNp}(a).

When comparing simulations with different initial semi-major axes in Fig.~\ref{fig: Rp_Ecc_single_combined}(a) for the same $M_0$, we find that the mean eccentricity is systematically larger for larger $a_{\mathrm{in},0}$. This trend arises from the higher Safronov number in those systems.
We discuss this effect in more detail later in the paper.

Fig.~\ref{fig: Rp_Ecc_single_combined}(a) shows that, overall, the mean eccentricity increases with planetary radius, regardless of composition. For rocky planets, the mean eccentricity is typically highest near the radius valley, whereas for icy planets it is highest around $\sim 3$–$4 R_\oplus$.
Fig.~\ref{fig: Rp_Ecc_single_combined}(b) shows that rocky and icy planets are clustered near $\sim1.4 R_\oplus$ and $\sim2.4 R_\oplus$, respectively. When rocky and icy planets are present in the overall population, the contrast between their radii naturally produces a radius valley \citep[e.g.,][]{Venturini+2020, Izidoro+2022, Burn+2024, Shibata+2025a}.
Figures~\ref{fig: Rp_Ecc_single_combined}(a)-(b) suggest that considering both populations together appears to be essential for matching the observations.

One might argue that combining planetary systems composed exclusively of rocky planets and exclusively of icy planets (such as those in S05 and S09) could produce a modest eccentricity enhancement near $\sim2R_\oplus$. However, as shown in Fig.~\ref{fig: Rp_Ecc_single_psNp}, the rocky planets responsible for the elevated eccentricities near the radius valley arise mainly in simulations with $N_0=18$ and $22$. Requiring systems to form on the order of twenty rocky protoplanets may appear extreme. Indeed, such large numbers of rocky protoplanets are not supported by our previous formation models \citep{Shibata+2025a, Shibata+2025b}. If we restrict our analysis to simulations with $N_0\lesssim10$, the resulting $\langle e\rangle$–$R$ curves become essentially flat, with mean eccentricities around the radius valley $\lesssim0.05$, which is inconsistent with the observed values. This indicates that an additional mechanism must be at work to enhance orbital instability and selectively raise the eccentricities of planets near the radius valley.

However, in these simulations all planets are assumed to start with identical masses and compositions. As we show later, the situation changes once a dichotomy in the mass distribution is allowed.

\section{Two-population systems}\label{sec: two_population}



In this section, we present additional N-body simulations that include interior rocky and external icy protoplanets. We assume that the rocky and icy protoplanets follow different formation pathways, leading to different characteristic masses. The primary goal of the simulations in this section is to isolate the dynamical effects of a mass contrast between the rocky and icy populations. Their compositions (and hence core densities) are not the main focus of this work. The full set of parameters used in the two-population simulations is summarized in Table~\ref{tab: parameters_for_two_pop}.

\begin{table}[h]
    \centering
    \begin{tabular}{|c|c|c|c|c|c|}
    \hline
     Sim. ID & $a_\mathrm{in,0}$ & $N_\mathrm{0}$ & $M_\mathrm{0}$ & $N_\mathrm{ext,0}$ & $M_\mathrm{ext,0}$ \\ 
    \hline
    \hline
    T00 & $0.05$ au & 10 & $1.0 M_\oplus$ & [4--10] &  $4M_\oplus$       \\ 
    T01 & $0.05$ au & 5 & $2.0 M_\oplus$ & [4--10] &  $8M_\oplus$      \\ 
    T02 & $0.05$ au & 10 & [0.5--2] $M_\oplus$ & 4 &  $4M_\oplus$       \\ 
    T03 & $0.05$ au & 10 & [0.5--2] $M_\oplus$ & 6 &  $4M_\oplus$       \\ 
    T04 & $0.05$ au & 10 & [0.5--2] $M_\oplus$ & 4 &  $6M_\oplus$       \\ 
    \hline

    \end{tabular}
    \caption{Parameters defining the configurations of two-population protoplanetary systems.}
    \label{tab: parameters_for_two_pop}
\end{table}
\begin{figure}
    \centering
    \includegraphics[width=\linewidth]{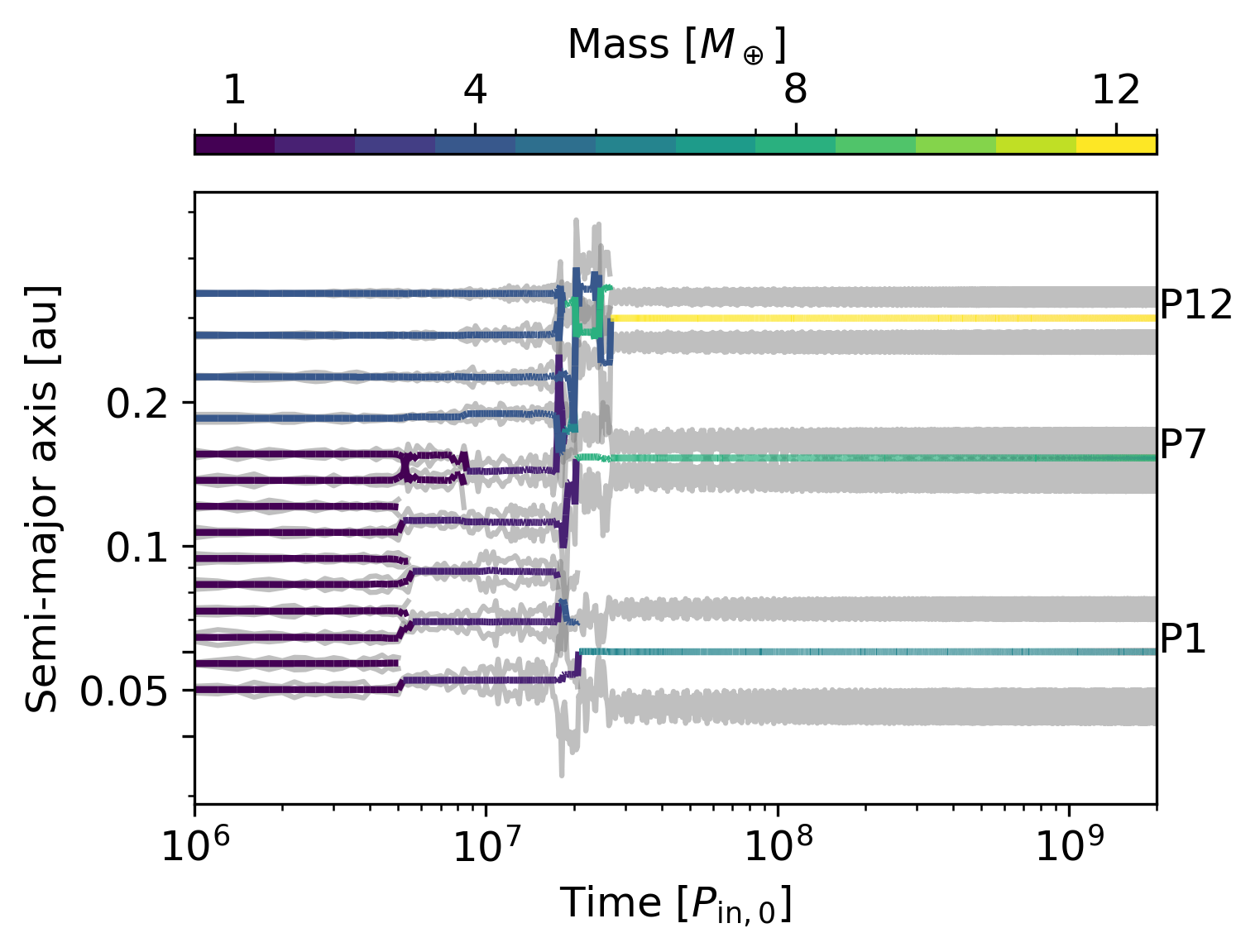}
    \caption{
    Same as Fi.g~\ref{fig: orb_single}, but for protoplanets in two-population systems. Here we show the case of $N_\mathrm{0}=10$, $M_\mathrm{0}=1M_\oplus$, $N_\mathrm{ext,0}=4$, $M_\mathrm{ext,0}=4M_\oplus$, and $a_\mathrm{in,0}=0.05$ au.
    }
    \label{fig: orb_twopop}
\end{figure}

Figure~\ref{fig: orb_twopop} shows a representative case of the orbital evolution in a two-population system. In this example, the system begins with $N_{0}=10$ rocky protoplanets of mass $M_{0}=1M_\oplus$ and $N_{\mathrm{ext},0}=4$ icy protoplanets of mass $M_{\mathrm{ext},0}=4M_\oplus$. Compared with the single-population case without external icy planets shown in Fig.~\ref{fig: orb_single}, the system undergoes an overall more violent orbital instability. On average, in this set of simulations about 2.2 rocky planets survive at the end, whereas 4.6 rocky planets survive in otherwise identical simulatios without external icy protoplanets (single population systems).

\begin{figure}
    \centering
    \includegraphics[width=\linewidth]{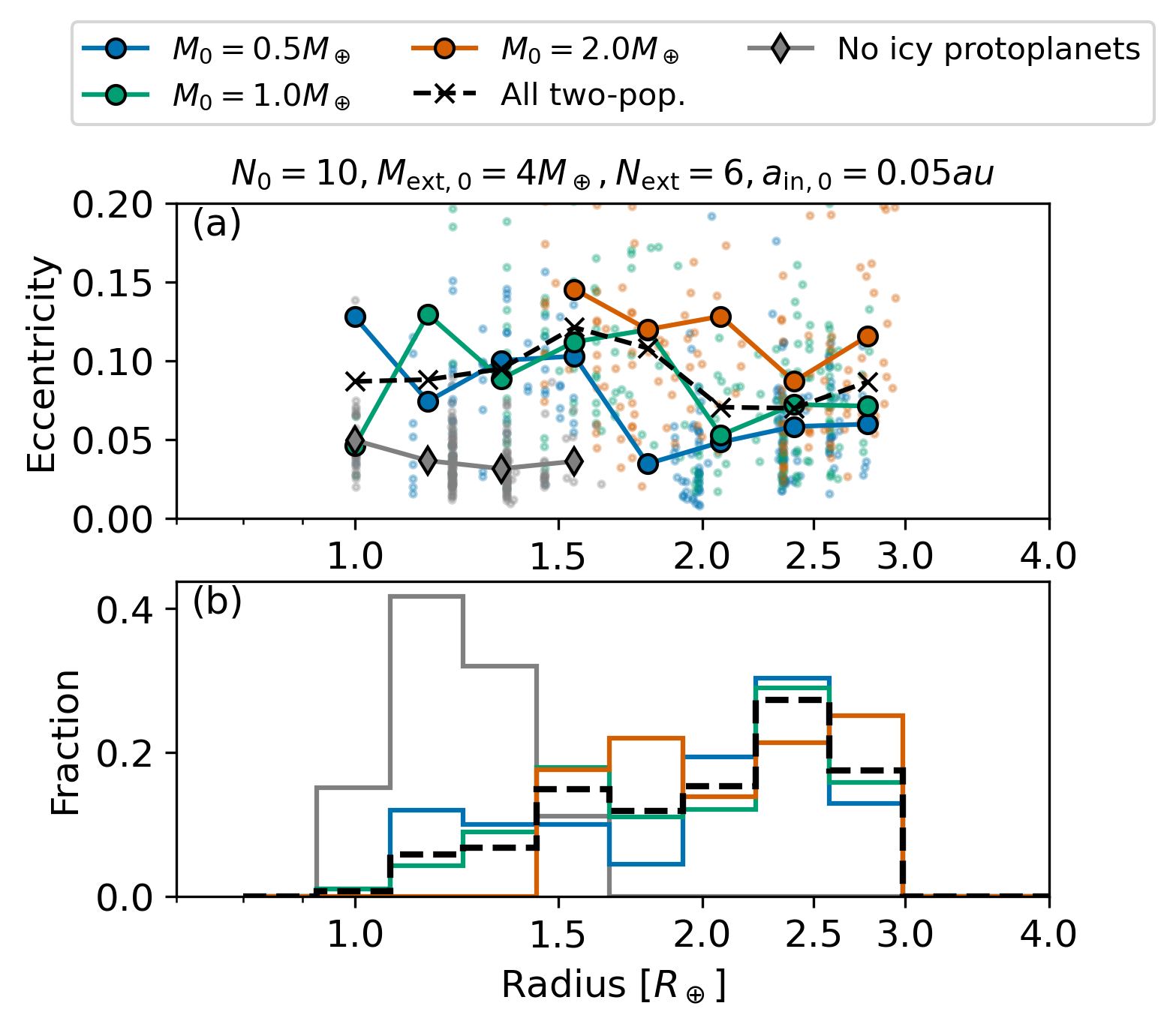}
    \caption{
    {\bf Panel-(a)}: mean eccentricity of planets as a function of their radius in two-population systems. The solid lines show the mean eccentricity of rocky and icy planets in each bin. The different colors correspond to the cases with different initial mass of inner rocky protoplanets $M_\mathrm{0}$, as shown in the top legend box. 
    Here, we show the cases where $a_\mathrm{in,0}=0.05$ au, $N_\mathrm{0}=10$ and $M_\mathrm{ext,0}=4M_\oplus$.
    {\bf Panel-(b)}: distribution of the adjusted radius of planets. The bin size is the same as in panel (a).
    }
    \label{fig: Rp_ecc_twopop_psMf}
\end{figure}

Figure~\ref{fig: Rp_ecc_twopop_psMf}(a) shows the mean eccentricity as a function of planetary radius for systems with $N_0=10$ inner rocky protoplanets and $N_{\rm ext,0}=6$ outer icy protoplanets. In all cases,  icy protoplanets have initial masses of $M_{\rm ext,0}=4M_\oplus$, while the different colors correspond to different values of $M_0$ for  rocky protoplanets, as indicated in the figure. Because the initial number of rocky (10) and icy (6) planets is held fixed, and the only free parameter is the rocky-planet mass $M_0$, this setup allows us to attempt to isolate the effect of the initial mass ratio ($M_{\rm ext,0}/M_0$) between the two populations..

In the simulations with $M_0=0.5 M_\oplus$,  rocky planets  ($\lesssim 1.5 R_\oplus$) have higher eccentricities than icy planets ($\gtrsim 2.0 R_\oplus$). This difference becomes less pronounced as the initial mass ratio $M_{\rm ext,0}/M_0$ decreases (see orange line). This can be understood by recalling that during gravitational scattering, random kinetic energy associated with orbital eccentricity is transferred from the more massive icy bodies to the less massive rocky ones via energy equipartition. As a result, icy bodies tend to have eccentricities damped while the rocky planets are excited. This exchange becomes progressively more efficient as the mass ratio between interacting planets increases and correspondingly weaker when the protoplanets have comparable masses.

As a sanity check for our interpretation, we calculate the mean eccentricity of all rocky planets $\langle e_\mathrm{rocky} \rangle$ and icy planets $\langle e_\mathrm{icy} \rangle$, and obtain the mean eccentricity ratio $\langle e_\mathrm{rocky} \rangle/ \langle e_\mathrm{icy} \rangle = 2.00, 1.73$, and $1.46$ for $M_0=0.5 M_\oplus$, $1.0 M_\oplus$, and $2.0 M_\oplus$, respectively. The mean eccentricity ratio increases as the initial mass ratio becomes larger. If icy protoplanets are about eight times more massive than rocky planets, the rocky super-Earths develop eccentricities that are roughly twice those of the icy mini-Neptunes. 

Figure~\ref{fig: Rp_ecc_twopop_psMf}(a) reveals another notable feature: 
the presence of external icy protoplanets triggers strong orbital instabilities, 
leading  rocky protoplanets to experience more giant impacts and grow to larger sizes. 
This trend is evident when comparing the blue and gray curves in 
Figure~\ref{fig: Rp_ecc_twopop_psMf}(a). In systems without external icy bodies, rocky planets alone tend to retain lower eccentricities and remain relatively small. In contrast, when icy protoplanets are present, numerous rocky planets acquire high eccentricities and grow to radii of $R \simeq 1.5$--$2\,R_\oplus$, clustering near the observed radius valley. This result does not imply that mixing rocky and icy planets in the same system is inherently inconsistent with the observations; rather, it suggests that this specific two-population systems selectively form eccentric planets in the radius valley.



\begin{figure}
    \centering
    \includegraphics[width=\linewidth]{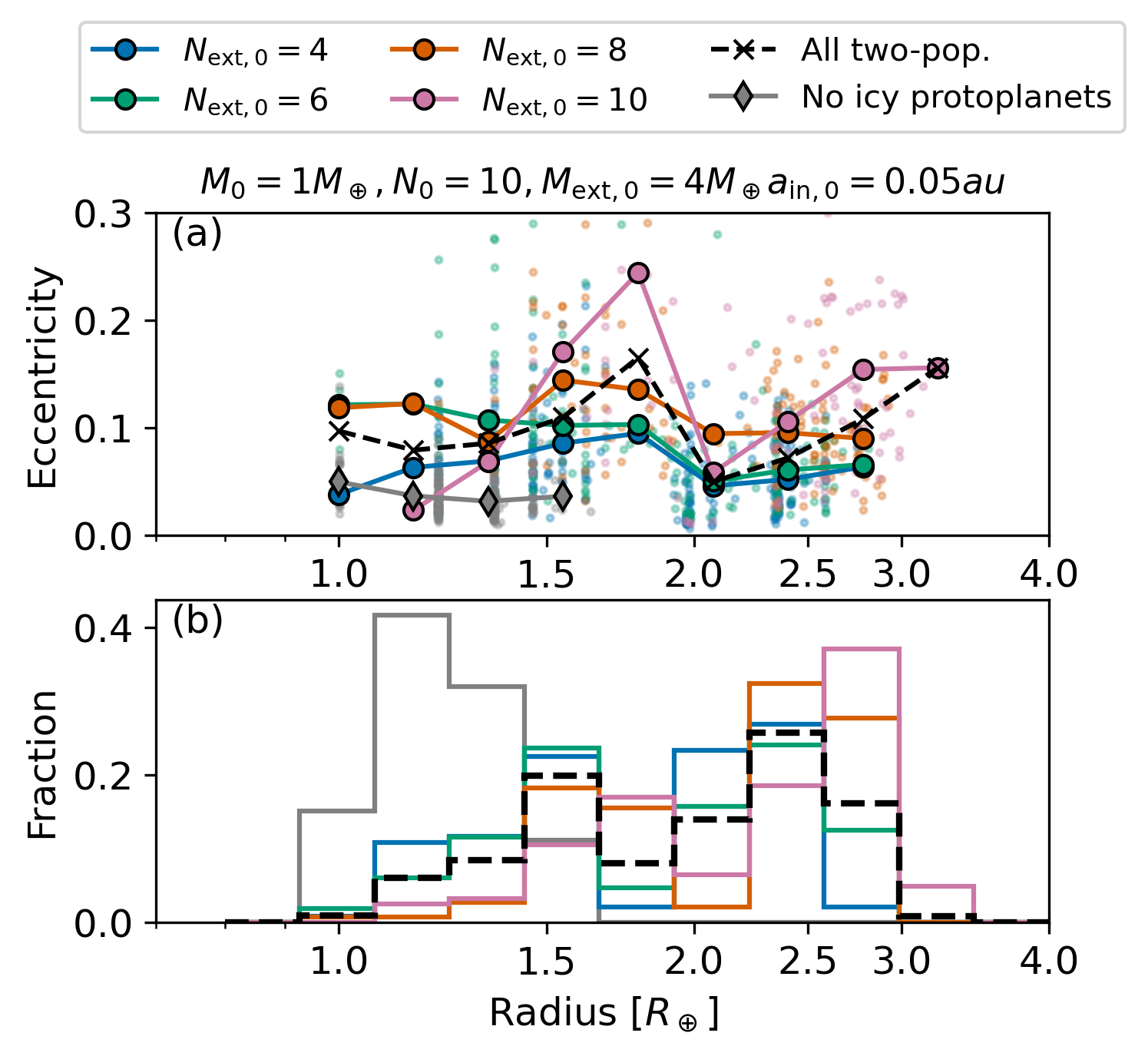}
    \caption{
    Same as Fig.~\ref{fig: Rp_ecc_twopop_psMf}, but with different parameter sets. The different colors correspond to the cases with different numbers of external icy protoplanets $N_\mathrm{ext,0}$. The other parameters are set to $a_\mathrm{in,0}=0.05$ au, $N_\mathrm{0}=10$ and $M_\mathrm{0}=1M_\oplus$.
    }
    \label{fig: Rp_ecc_twopop_psNp}
\end{figure}

We now vary the number of icy protoplanets while keeping all other relevant parameters fixed. 
Figure~\ref{fig: Rp_ecc_twopop_psNp}(a) shows the mean eccentricity as a function of planetary radius 
for systems with $N_0=10$ inner rocky protoplanets of initial mass $M_0=1M_{\oplus}$ 
while we vary the initial number of external icy protoplanets $N_{\mathrm{ext},0}$. 
In all cases,  icy protoplanets have initial masses of $M_{\mathrm{ext},0}=4M_{\oplus}$.

As $N_{\mathrm{ext},0}$ increases, the total mass in icy bodies becomes larger and dynamical instabilities 
among the icy protoplanets themselves become stronger. Consequently, although the rocky planets also tend to 
be more strongly excited, the eccentricity ratio 
$\langle e_{\mathrm{rocky}}\rangle/\langle e_{\mathrm{icy}}\rangle$ does not increase correspondingly 
because the icy planets likewise experience enhanced eccentricity excitation. In fact, the outermost icy protoplanets predominantly scatter each other while only those located closest 
to the rocky population effectively participate in driving up the rocky planets' eccentricities. 
An excess of large icy planets in the system is expected to drive higher eccentricities for planets at larger radii (e.g., $>3$–$4R_{\oplus}$).


\section{Combining Single and Two-population systems}\label{sec: compare_obs}

\subsection{Simulations}


Planet formation models predict a range of planetary architectures: some disks yield predominantly rocky planets, others produce icy planets that migrate inward, and still others form mixed rocky–icy systems~\citep{izidoroetal21,Izidoro+2022,Shibata+2025a,Shibata+2025b}. Motivated by this diversity, we now combine all our different scenarios and assess their implications for the observed radius valley and eccentricity distribution.

\begin{figure}
    \centering
    \includegraphics[width=\linewidth]{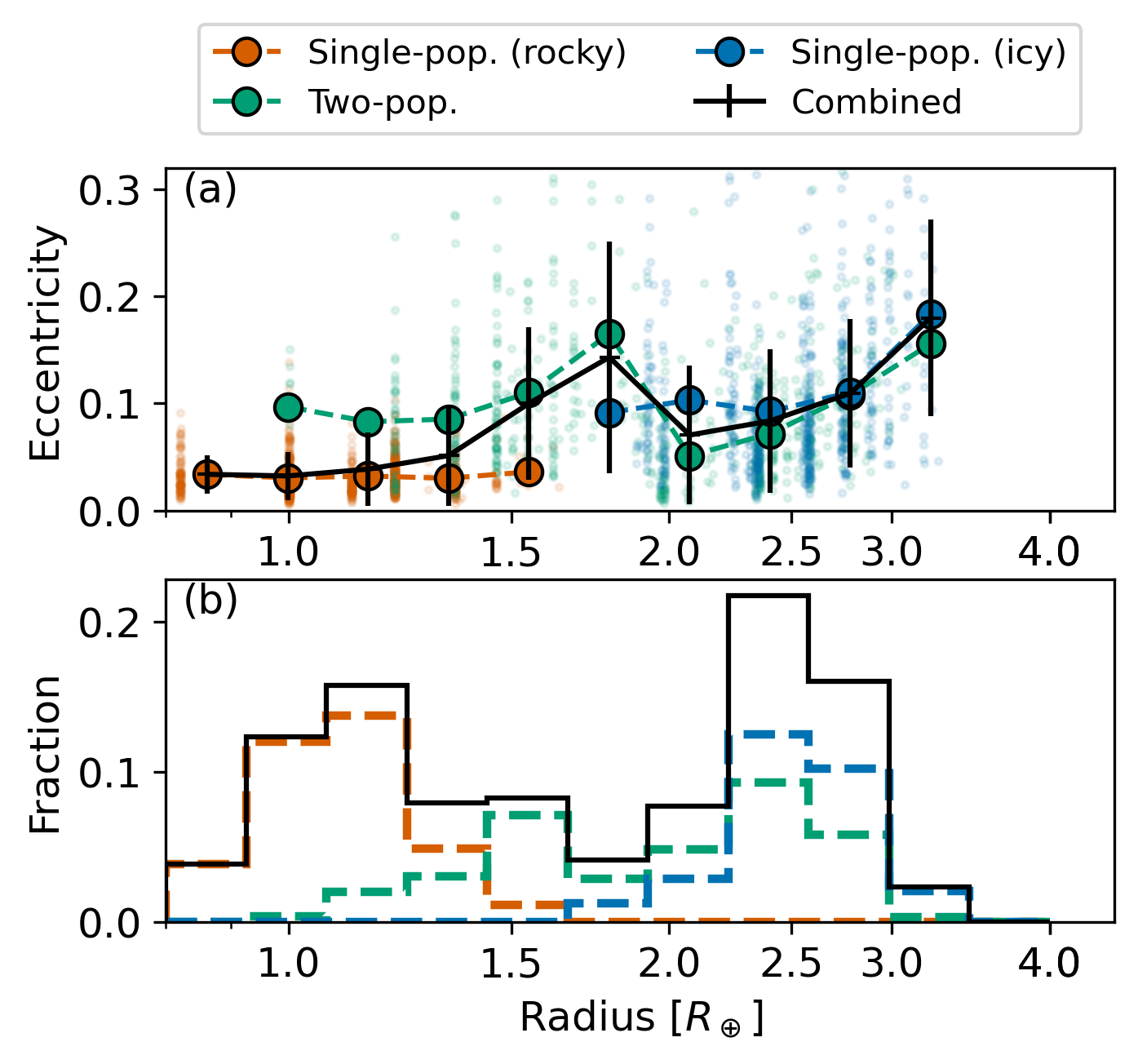}
    \caption{
    Same as Fig.~\ref{fig: Rp_ecc_twopop_psMf}, but mixed with single- and two-population systems. The orange and blue lines show the single-population cases of rocky (S02 and S04) and icy protoplanets (S07). The green dashed line shows the case of two-population case (T00). The black solid line shows the result mixed with the single- and two-population systems. From the rocky single-population systems, we remove the simulations of $N_0>10$.
    }
    \label{fig: Rp_ecc_mix_all}
\end{figure}


In this section, we combine the results of our simulations that include rocky planets only (S02 and S04), 
icy planets only (S07), and mixed rocky–icy systems (T00). 
We include all parameter combinations explored in these scenarios, 
except for the rocky single-population cases with $N_0>10$, 
which we exclude because forming systems with more than 10 rocky protoplanets is difficult within current 
formation models~\citep{Shibata+2025a, Shibata+2025b}.  Figure~\ref{fig: Rp_ecc_mix_all} shows the distribution of planetary radius and mean eccentricity for  these mixed populations.


It is interesting to note in Figure~\ref{fig: Rp_ecc_mix_all}(a) that the single-population rocky systems 
(orange) typically produce planets with small eccentricities and radii ranging from $1$ to 
$1.5\,R_{\oplus}$. In contrast, the single-population icy systems tend to produce planets whose 
eccentricities increase with size. The two-population cases (green) correspond to the combined set of 
outcomes shown in Figure~\ref{fig: Rp_ecc_twopop_psNp}. 


Taken together, our results suggest that neither purely rocky systems nor purely icy systems alone can reproduce the observed structure of the radius valley and its associated eccentricity distribution. Two-population systems provide a more promising pathway, but even these do not fully capture the observed trend across the entire range of planetary radii. A more consistent match with observations likely requires a mixture of planetary-system types: some that are purely rocky, some that are purely icy, and some that contain both rocky and icy planets. In the next section, we account for observational selection effects applied to our simulated planetary systems and directly compare the resulting synthetic population to observations.

\subsection{Comparing our results and observations}
To enable a meaningful comparison between the simulated and observed super-Earths and mini-Neptunes, we simulate transit observations for each planetary system in our ensemble, following the method developed by \citet{Izidoro+2017}. In this section, we compare the resulting synthetic eccentricity distribution with the observed distribution reported by \citet{Gilbert+2025}.

\begin{figure}
    \centering
    \includegraphics[width=\linewidth]{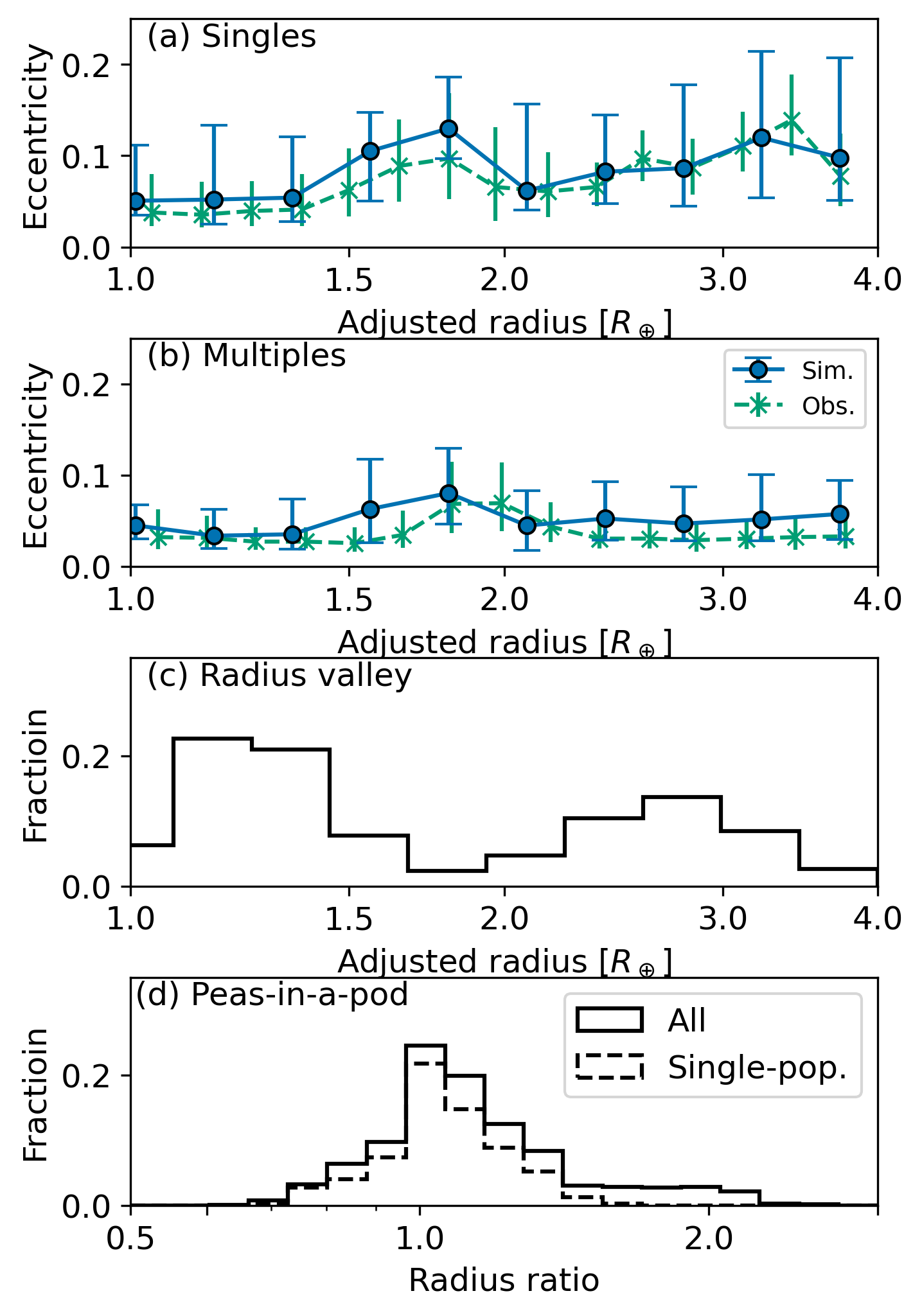}
    \caption{
    {\bf Panel-(a)} and {\bf (b)}: Comparison of our numerical model to observations. The solid blue lines show $\langle e \rangle$-$R$ distribution obtained in our numerical results, including observational biases. The green dashed lines show that of observational data. The planets are grouped to (a) single-observed and (b) multiple-observed planets. The observational data were extracted from Fig. 4 in \citet{Gilbert+2025} through visual inspection using PlotDigitizer ({\rm https://plotdigitizer.com/app}).
    {\bf Panel-(c)}: Radius distribution of our numerical model.
    {\bf Panel-(d)}: Peas-in-a-pod, or radius radio distribution, of our numerical model. The dashed line shows the contribution from the single-population systems. The numerical data used for this figure are listed in Tab.~\ref{tab: parameters_for_mixing_obs}.
    }
    \label{fig: Rp_ecc_mix_Multiple}
\end{figure}

\begin{table}[h]
    \centering
    \begin{tabular}{|c|c|}
    \hline
    Type of planetary system & [$a_\mathrm{in,0}$, $N_\mathrm{0}$, $M_\mathrm{0}$, $N_\mathrm{ext,0}$, $M_\mathrm{ext,0}$]\\ 
    \hline
    \hline
    single population: rocky  & [$0.05$ au, 6, $1.0 M_\oplus$, -, -] \\
                        & [$0.05$ au, 10, $1.0 M_\oplus$, -, -] \\
    \hline
    two population: rocky \& icy   & [$0.05$ au, 10, $0.5 M_\oplus$, 4, $4M_\oplus$] \\
                        & [$0.05$ au, 10, $1.0 M_\oplus$, 4, $6M_\oplus$] \\
    \hline
    single population: icy & [$0.1$ au, 6, $4.0 M_\oplus$, -, -] \\
                     & [$0.1$ au, 6, $8.0 M_\oplus$, -, -] \\
                     & [$0.1$ au, 8, $8.0 M_\oplus$, -, -] \\
    \hline
    \end{tabular}
    \caption{Simulations used for Fig.~\ref{fig: Rp_ecc_mix_Multiple}}
    \label{tab: parameters_for_mixing_obs}
\end{table}

When comparing simulations (including observational bias) and observation estimates we use the ``adjusted radius'' (rather than the real radius) as introduced in \citet{Ho+2023}. It is defined as
\begin{equation}
    R_\mathrm{adj} = R \left( \frac{P}{10~\mathrm{day}} \right)^{m},
    \label{eq: Rpadj}
\end{equation}
with $m=0.10$ \citep{Petigura+2022}. 
In addition, we show the median eccentricities instead of the mean values, and we use the 16th and 84th 
percentiles as error bars, following \citet{Gilbert+2025}.

We combine three types of planetary systems and seven parameter sets, as summarized in Table~\ref{tab: parameters_for_mixing_obs}.  Motivated by the findings of \citet{Gilbert+2025}, who reported distinct $\langle e \rangle$--$R_\mathrm{adj}$ distributions for single- and multiple-transiting systems, we divide our results into two groups: systems with only one transiting planet and those with two or more. The blue curves in Figure~\ref{fig: Rp_ecc_mix_Multiple} show the median eccentricities of the super-Earths and mini-Neptunes from our synthetic observations. Panels~(a) and (b) show singles and multiples, respectively. The green dashed line show the eccentricities reported by  \citet{Gilbert+2025}. Figure~\ref{fig: Rp_ecc_mix_Multiple}(c) shows the radius distribution of our simulated observations population, where a clear valley is observed at about 1.8$R_{\oplus}$, which is broadly consistent with observations. Panel~(d) shows the distribution of radius ratios for adjacent planet pairs, highlighting the characteristic ``peas-in-a-pod'' pattern reported by \citet{Weiss+2018}.

\citet{Gilbert+2025} found three important features in the observed eccentricity distribution: (i) the eccentricity distribution peaks around the radius valley among both single and multiple planets, (ii) the peak is smaller among multiple planets than among single planets, and (iii) the eccentricity increases toward larger radii only for single mini-Neptunes. Our model successfully reproduces these features observed in exoplanetary systems.

We note that the parameter sets used in Figure~\ref{fig: Rp_ecc_mix_Multiple} were selected somewhat arbitrarily. Varying the combination of parameters alters the resulting $\langle e \rangle$--$R_\mathrm{adj}$ trend, in some cases producing noticeable deviations from the observations. For example, the initial mass fraction of the inner rocky and outer icy protoplanets directly affects the prominence of the peak near the radius valley, since the degree of energy equipartition depends on this mass fraction. Additionally, we adopted a discontinuous mass distribution for the protoplanets (e.g. 1 and 2$M_{\oplus}$), which may result in larger error bars than observed; a smooth distribution of masses, such as a Gaussian, may be more appropriate. 

It is important to emphasize that our aim here is not to obtain a perfect match to the observations, but rather to provide broad physical insight into the conditions required to reproduce the observed features. Future work exploring a wider and finer parameter space will be valuable for placing stronger constraints on the structure of primordial planetary systems.


\section{System Safronov number and the architecture of initial planetary systems}\label{sec: Safronov}

\begin{figure}
    \centering
    \includegraphics[width=1.0\linewidth]{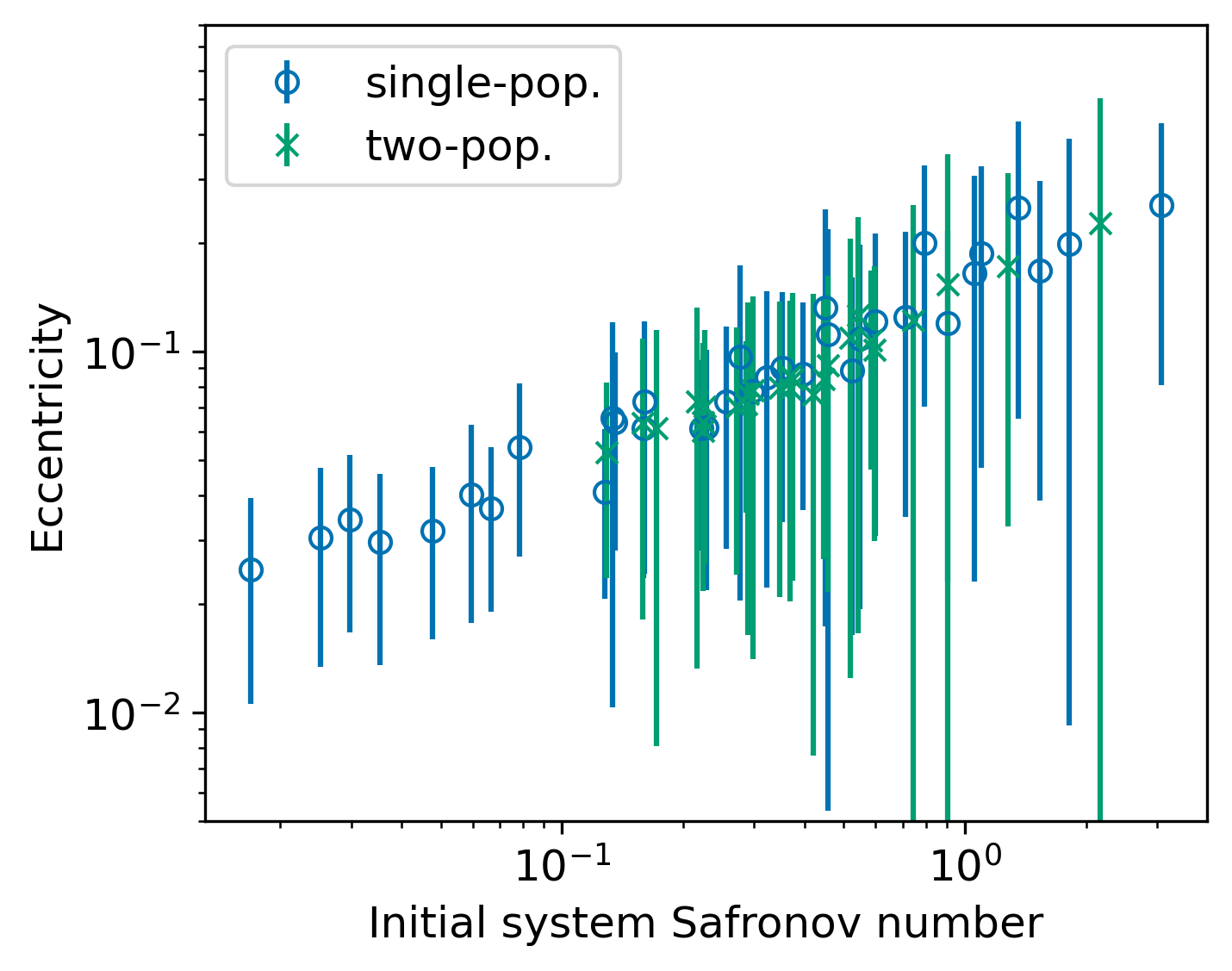}
    \caption{
    Mean eccentricity as a function of the initial system Safronov number. The mean eccentricity is calculated for each parameter study, and the error bars show the standard deviation. The circles and crosses show the simulations of single- and two-population systems.
    }
    \label{fig: Ecc_Theta}
\end{figure}

Before proceeding, it is worth noting that the analysis that follows is not directly tied to the preceding population-level comparisons. Instead, its purpose is to provide a more quantitative and physically grounded interpretation of how the initial architecture of a planetary system influences its final eccentricity, post instability. By examining this connection through the lens of the Safronov number, we aim to identify the key physical parameters controlling eccentricity excitation, rather than relying solely on phenomenological or empirical trends.

Our simulations show that the level of eccentricity excitation within the radius valley depends sensitively on the masses, numbers, and semi-major axes of the protoplanets. Recently, \citet{Kokubo+2025} reported that the characteristic final eccentricity of planetary systems scales not with the Hill radius, as traditionally assumed, but instead with a length scale related to the Safronov number. This finding is somewhat surprising, as it shifts a commonly held expectation in the field. The Safronov number $\Theta$ is defined as
\begin{align}
    \Theta = \frac{{v_\mathrm{esc}}^2}{{v_\mathrm{K}}^2} = \frac{M_\mathrm{p} a_\mathrm{p}}{M_\mathrm{s} R_\mathrm{p}}, \label{eq: system_Safronov}
\end{align}
where $M_\mathrm{p}, R_\mathrm{p}$, and $a_\mathrm{p}$ are the mass, radius, and semi-major axis of the protoplanet. We confirm that their scaling laws hold in our numerical simulations (see Appendix~\ref{app: scaling_laws}). While their analysis primarily considered the final orbital architectures of planetary systems, our results allow us to extend this perspective by examining how the initial orbital configuration influences the final eccentricity. To this end, we further analyze our simulations using the Safronov number.

The Safronov number is related to the scattering-to-collision ratio \citep[e.g.][]{Shibata+2024}. In order to increase the protoplanet's eccentricity, the scattering velocity of close encounter, $v_\mathrm{scat}=\sqrt{\mathcal{G} M_\mathrm{p} /b}$, needs to be larger than the epicyclic velocity of the protoplanet $v_\mathrm{scat}\gtrsim e_\mathrm{p} v_\mathrm{K}$, where $\mathcal{G}$ is the gravitational constant, $b$ is the distance of the close encounter, and $e_\mathrm{p}$ is the eccentricity of the protoplanet. This equation gives a maximum radius for scattering protoplanets $R_\mathrm{scat}$ as
\begin{align}
    b = \frac{1}{e_\mathrm{p}^2} \frac{M_\mathrm{p}}{M_\mathrm{s}} a_\mathrm{p} 
    = \frac{\Theta}{e_\mathrm{p}^2} R_\mathrm{p}
    \equiv R_\mathrm{scat}.
\end{align}
The collision radius is given by
\begin{align}
    R_\mathrm{col} = R_\mathrm{p} \left( 1 + \frac{\Theta}{e_\mathrm{p}^2} \right)^{1/2},
\end{align}
where the bracket is the gravitational focusing factor, and we assume the relative velocity is given by $e_\mathrm{p} v_\mathrm{K}$.
Therefore, the scattering-to-collision ratio is given by
\begin{align}
    \lambda = \frac{\pi {R_\mathrm{scat}}^2}{\pi {R_\mathrm{col}}^2} = \frac{\left(\Theta/e_\mathrm{p}^2\right)^2}{1+\Theta/e_\mathrm{p}^2}. \label{eq: lambda}
\end{align}
$\lambda$ represents the typical number of close encounters before the protoplanet collides with other protoplanets. For each planetary system, the expected total number of close encounters is given by
\begin{align}
    \Lambda = \sum_{i=1}^{N} \lambda_i.
\end{align}

If $\Theta\gg{e_\mathrm{p}}^2$, $\lambda\propto\Theta$ and  
\begin{align}
    \Lambda \propto \sum_{i=1}^{N} \Theta_i := \Theta_\mathrm{sys},
\end{align}
where we newly define the system Safronov number $\Theta_\mathrm{sys}$. 

Although $\Theta_\mathrm{sys}$ evolves as protoplanets collide and grow, it scales as $\Theta_\mathrm{sys}\propto M_\mathrm{p}^{-1/3}$ under our assumptions of $N M_\mathrm{p}=\mathrm{const.}$ and a constant bulk density. Consequently, the initial architecture of the planetary system has the dominant impact on the total number of close encounters the system can experience.

Figure~\ref{fig: Ecc_Theta} shows the final mean eccentricity of planets, $\langle e \rangle$, as a function of the initial system Safronov number, $\Theta_\mathrm{sys,0}$. We find that the final mean eccentricity correlates well with the initial system Safronov number. Physically, the mean eccentricity reflects the number of close encounters experienced during the orbital instability phase. Because energy equipartition acts differently on rocky and icy planets, two-population systems exhibit a broader spread in mean eccentricities (see error-bars) than single-population systems.

Note that
\begin{align}    
    \frac{\Theta}{e_\mathrm{p}^2} = 71 & \left( \frac{e_\mathrm{p}}{0.01} \right)^{-2}
    \left( \frac{M_\mathrm{p}}{1 M_\oplus} \right)
    \left( \frac{M_\mathrm{s}}{1 M_\odot} \right)^{-1} \nonumber \\
    & \left( \frac{a_\mathrm{p}}{0.1 \mathrm{au}} \right)
    \left( \frac{R_\mathrm{p}}{1 R_\oplus} \right)^{-1}.
\end{align}
Therefore, $\Theta\ll{e_\mathrm{p}}^2$ when $e_\mathrm{p} \gtrsim 0.1$. This means that scattering occurs less frequently as the planetary system becomes more dynamically excited. 


In our simulations, the initial system Safronov number $\Theta_\mathrm{sys,0}$ is given as
\begin{align}
    \Theta_\mathrm{sys,0} &= \frac{M_\mathrm{0} a_\mathrm{in,0}}{M_\odot R_\mathrm{0}} \sum_{i=1}^{N_\mathrm{0}} \left( \frac{2 +k_\mathrm{orb} h}{2 -k_\mathrm{orb} h} \right)^{i-1}, \\
    &= \frac{M_\mathrm{0} a_\mathrm{in,0}}{M_\odot R_\mathrm{0}} \frac{r^{N_\mathrm{0}}-1}{r-1}, \label{eq: sysTheta_model}
\end{align}
with
\begin{align}
    r = \frac{2 +k_\mathrm{orb} h}{2 -k_\mathrm{orb} h}.
\end{align}
Equation~\ref{eq: sysTheta_model} shows that for a fixed total mass of protoplanets $M_0 N_0$, larger $N_0$ with smaller $M_0$ results in a higher $\Theta_\mathrm{sys,0}$. This result is consistent with the results in Fig.~\ref{fig: Rp_Ecc_single_psMp} and Fig.~\ref{fig: Rp_Ecc_single_psNp}.

\citet{Kokubo+2025} found that the final eccentricities of planetary systems scale as $e \propto \Theta^{1/2}$. This scaling may arise from the random-walk nature of eccentricity excitation during close encounters. Each encounter between protoplanets of comparable masses can either increase or decrease eccentricity, and the cumulative effect resembles a random walk in $e$. Because the system Safronov number is related to the expected number of close encounters, one can heuristically write $e \propto {\Theta_\mathrm{sys,0}}^{1/2}$.

From our simulations, the best-fit power-law relationship between $e$ and the initial system Safronov number is $e \propto {\Theta_\mathrm{sys,0}}^{0.39}$, which is slightly shallower than the $1/2$ exponent (see Appendix~\ref{app: fitting_Ecc_Theta}). Physically, this weaker scaling likely arises because the probability of close encounters decreases as eccentricities grow (see Eq.~\ref{eq: lambda}). As the orbits become more excited, the efficiency of further eccentricity pumping diminishes, reducing the strength of the scaling between $e$ and $\Theta_\mathrm{sys,0}$.

\section{Caveats and discussions}\label{sec: caveats}

In this work, we study the dynamical evolution of compact planetary systems starting from configurations where adjacents planets are initially separated by $10$ mutual Hill radii. This setup ensures that most systems undergo orbital instability and experience strong planet--planet scattering and collisions. In this regime, our simulations show that mixing between rocky and icy planet populations (with distinct mass distributions) is a key ingredient for producing enhanced eccentricities at radii corresponding to the radius valley. Of course, our results do not rule out alternative scenarios that may also be consistent with observations.

One may argue, for instance, that  a large fraction of planetary systems may form with wider initial orbital separations than that we assume in our simulations. If this is the case, they can perhaps undergo only gentle dynamical reorganization, potentially avoiding collisions. In such systems, eccentricities may remain low, and the radius distribution may be shaped primarily by photoevaporation or core-powered mass-loss processes, yielding two peaks near $\sim 1.4\,R_\oplus$ and $\sim 2.4\,R_\oplus$. If only a minority of systems experience violent instabilities similar to those in our compact initial conditions, it is possible that the eccentricity enhancement near the radius valley would still arise through the same mechanism identified here, by mixing these two populations. In this scenario, the higher eccentricity in the radius-valley would not necessarily require two intrinsically distinct rocky and icy planet populations.

A potential challenge for this scenario is accounting for the inferred 
eccentricities of single-transiting mini-Neptunes 
($2$--$4\,R_\oplus$; see Fig.~\ref{fig: Rp_ecc_mix_Multiple}a). 
Single-transiting mini-Neptunes appear to exhibit systematically higher 
eccentricities than planets in multiple-transiting systems. In our model, these eccentricities arise naturally because many apparent 
``singles'' are not intrinsically single-planet systems. Instead, they 
often host additional nearby companions that do not transit due to 
geometric bias, yet dynamically excite the observed planet's 
eccentricity and inclinations \citep{Izidoro+2017}.

If mini-Neptunes formed on primordially well-separated orbits and 
remained dynamically isolated, most single-transiting planets would be 
expected to be as dynamically cold as planets in multiple-transiting systems. The observed eccentricity excess among singles therefore 
supports the interpretation that many of them are members of  intrinsically multi-planet systems whose companions are observationally missed.

We also neglect the influence of very distant external perturbers (e.g., $\gg 1$~au) on the inner planets. Such companions can promote or sustain instabilities in compact inner systems \citep{bitschizidoro24,Ogihara+2025} and may therefore modify the eccentricity--radius relation. Quantifying this effect requires dedicated modeling that couples inner-system dynamics to the statistics of outer companions.

Overall, while our proposed formation pathway is consistent with the available observational constraints, it is important to test whether alternative formation channels can reproduce the same eccentricity--radius behavior. A systematic exploration of the eccentricity--radius evolution across different initial architectures, including the role of distant perturbers and dynamically cold configurations, is therefore a key direction for future work. This is particularly timely because upcoming missions such as \textit{PLATO} will better constrain the occurrence and architectures of outer planetary systems, enabling sharper tests of the origin of the radius valley and its associated eccentricity signatures.

\section{Summary and conclusion}\label{sec: conclusion} 

\begin{figure*}
    \centering
    \includegraphics[width=0.9\linewidth]{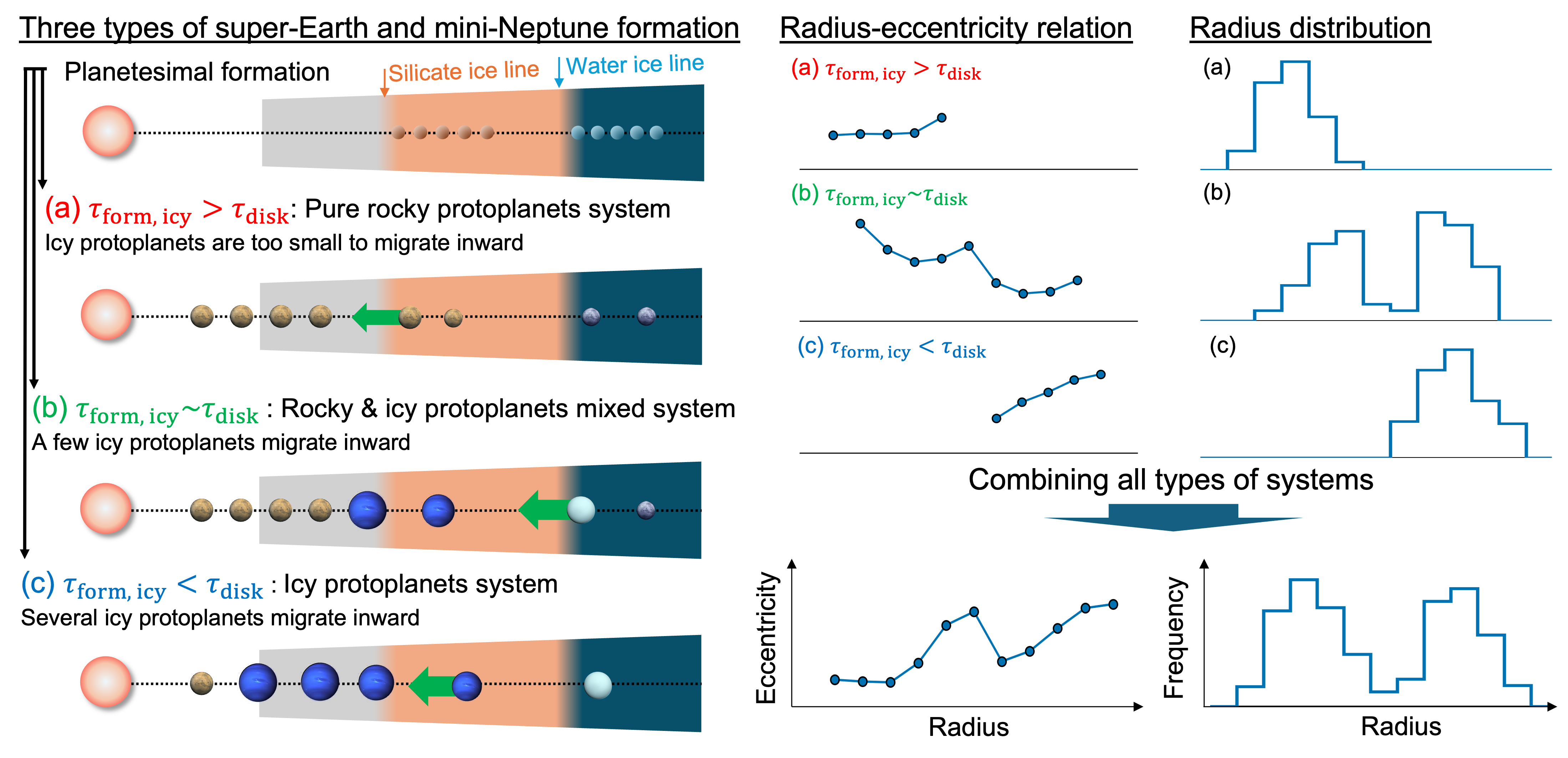}
    \caption{
    Schematic image of three types of protoplanetary systems and their contribution to the radius-eccentricity profile and radius valley.
    }
    \label{fig: three_systems}
\end{figure*}
In this study, we investigated the evolution of planetary radii and eccentricities in single- and two-population systems using a suite of $N$-body simulations. Our results indicate that the peak in the eccentricity–radius distribution arises from a mixture of planetary populations with different characteristic masses, implying that these populations likely follow distinct formation pathways. A plausible interpretation is that rocky and icy protoplanets, originally formed inside and outside the water iceline, migrated inward to close-in orbits. Interior to the iceline, the solid inventory is smaller and the pebble isolation mass is lower (if pebble accretion operates), favoring the formation of lower-mass rocky planets. Exterior to the iceline, a larger supply of solids and a higher pebble isolation mass promote the growth of more massive icy planets. These differences in characteristic masses help explain why the eccentricity–radius peak arises from a mixture of the two populations.

This framework can be naturally interpreted in the broader context of planet formation: the relative timing between icy protoplanet growth and disk dispersal determines whether close-in planets are rocky, mixed, or icy.  \citet{Shibata+2025a, Shibata+2025b} showed that the competition between the formation timescale of icy protoplanets, $\tau_\mathrm{form, icy}$, and the disk lifetime, $\tau_\mathrm{disk}$, determines whether a system becomes purely rocky, mixed rocky–icy, or predominantly icy. If the disk disperses before icy protoplanets can grow ($\tau_\mathrm{form, icy} > \tau_\mathrm{disk}$), the resulting system contains only rocky planets in close-in orbits (Fig.~\ref{fig: three_systems}--(a)). Conversely, if icy protoplanets form rapidly relative to the disk lifetime ($\tau_\mathrm{form, icy} < \tau_\mathrm{disk}$), many massive icy bodies migrate inward, disrupt resonant chains, and dominate the system, leaving few rocky planets behind (Fig.~\ref{fig: three_systems}--(c)). Mixed systems arise when $\tau_\mathrm{form, icy}$ is comparable to $\tau_\mathrm{disk}$ (Fig.~\ref{fig: three_systems}--(b)).

Taken together, our results support the existence of three characteristic classes of close-in planetary systems: (i) purely rocky systems, (ii) mixed rocky–icy systems, and (iii) predominantly icy systems. These outcomes emerge naturally from the interplay between protoplanet formation timescales, disk lifetimes, and inward migration, and they provide a coherent physical explanation for the diversity of observed eccentricity–radius distributions.

One may argue against our interpretation by invoking that the water-rich planets can result from atmosphere–interior interactions \citep{Kimura+2020, Miozzi+2025}, meaning that a water-rich composition alone does not necessarily imply formation beyond the iceline. However, this mechanism does not naturally produce the larger characteristic masses required to explain the observed eccentricity levels. In contrast, the eccentricity–radius relation is more directly tied to the underlying mass distribution of planets. 

Our simulations are intentionally idealized, enabling us to isolate the dominant physical processes that shape the eccentricity structure near the radius valley. One simplification is that planets of a given composition within the same system are assigned equal masses. While this assumption neglects possible intra-system mass dispersion, we do not expect it to qualitatively alter our results. Observationally, the ``peas-in-a-pod'' trend indicates that planets within a given system typically have comparable masses, supporting the view that modest variations in individual masses would preserve the main dynamical outcomes identified here.

Complementing our analysis, we also found that the mean eccentricity of planetary systems correlates well with the system Safronov number, defined as the sum of the Safronov numbers of the constituent planets. This reflects the fact that systems with larger characteristic masses experience stronger eccentricity excitation during close encounters. In addition, energy equipartition between lower- and higher-mass planets naturally leads to lower eccentricities for the larger-radius planets, producing the steep drop observed at the upper end of the radius distribution.

The results presented here broadly generalize and confirm the conclusions of \citet{Shibata+2025b}. Our simulations demonstrate how distinct formation pathways and system architectures lead to different levels of dynamical excitation and produce different signatures in the radius–eccentricity distribution, thereby clarifying the physical conditions required to reproduce the observed trends.

Overall, our results highlight the importance of considering both planetary composition and system architecture when modeling the origin of the radius valley. By linking the system Safronov number to dynamical excitation, we provide a unified physical framework for interpreting the eccentricity distribution of close-in planets. Future observations that constrain eccentricities and compositions across a wider range of planetary radii will be key to testing whether the diversity of planetary systems indeed arises from the interplay of rocky, icy, and mixed populations.


\begin{acknowledgments}
The authors are grateful for the constructive feedback
from the anonymous referee. This work was supported in part by the Big-Data Private-Cloud Research Cyberinfrastructure MRI-award funded by NSF under grant CNS-1338099 and by Rice University’s Center for Research Computing (CRC). The numerical computations were carried out on PC cluster at the Center for Computational Astrophysics, National Astronomical Observatory of Japan.
\end{acknowledgments}

\bibliography{sample701}{}
\bibliographystyle{aasjournalv7}

\newpage
\appendix

\section{Effects of the initial orbital separation}\label{sec: orbital_separation}

\begin{figure}[h]
    \centering
    \includegraphics[width=0.6\linewidth]{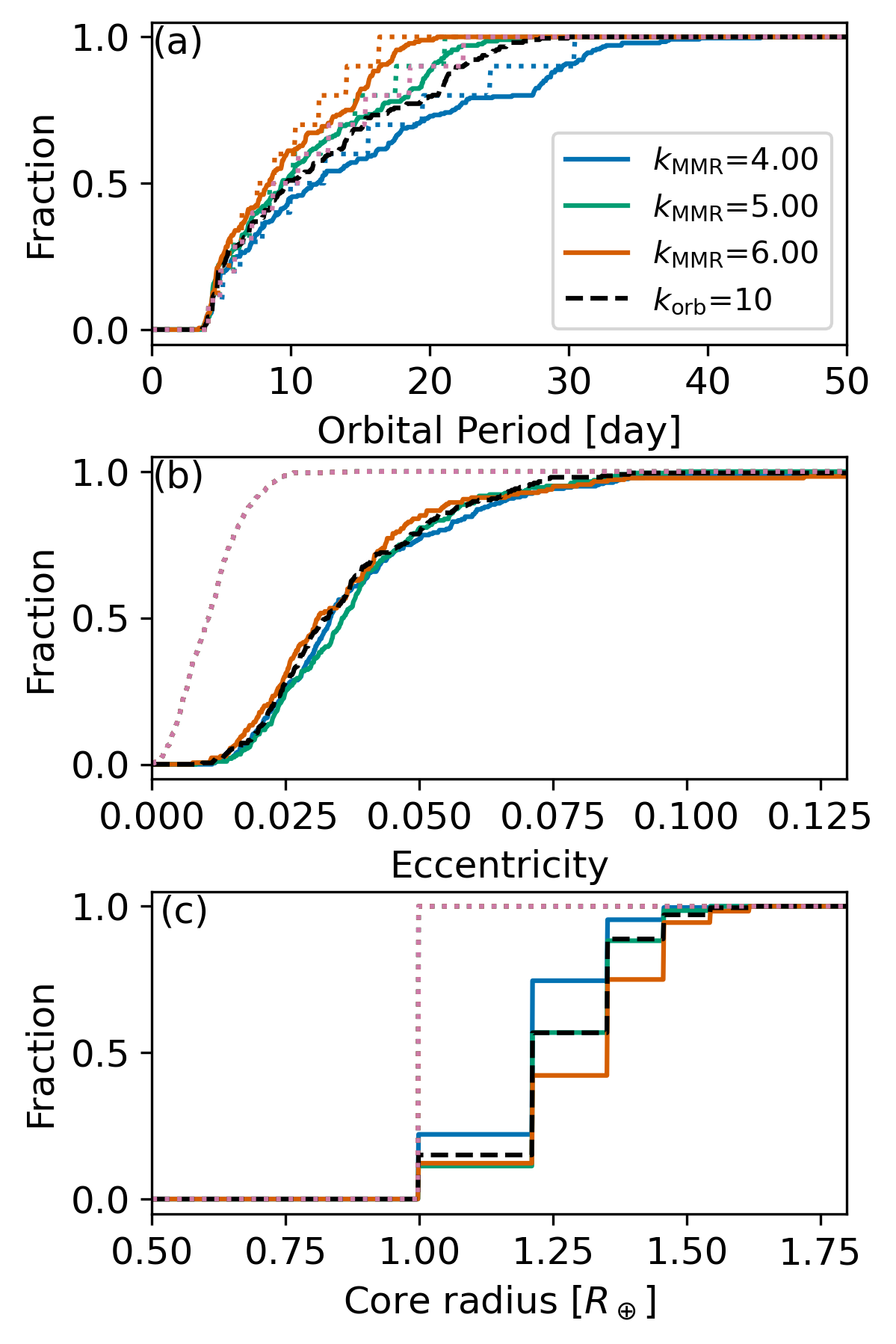}
    \caption{
    Cumulative distributions of planets' (a) orbital periods, (b) eccentricities, and (c) core radius. The different colors correspond to different orders of mean motion resonances $k_\mathrm{MMR}$, as shown in the legend box. The dashed black lines are cases where the protoplanets are separated by $k_\mathrm{orb}=10$ mutual Hill radii. The dashed lines show the initial distribution.
    }
    \label{fig: Rp_Ecc_psfecc0}
\end{figure}

In the simulations of the main paper, we fixed the initial orbital separation of protoplanets to $k_\mathrm{orb}=10$. However, within a self-cosistent framework of planet formation, protoplanets are expected to be trapped in mean-motion resonances, implying that the actual orbital separations may differ from those assumed in our nominal simulations. To clarify the effect of this assumption, we ran additional simulations in which the orbital spacing is determined by the resonance configuration. When protoplanets are trapped in $k_\mathrm{MMR}:k_\mathrm{MMR}+1$ mean-motion resonances, their initial semi-major axes are given by  
\begin{align}
    a_{i+1,0} = a_{i,0} \left(\frac{k_\mathrm{MMR}+1}{k_\mathrm{MMR}}\right)^{2/3}.
\end{align}

In \citet{Shibata+2025a}, most protoplanets are trapped in resonances with $k_\mathrm{MMR}=4,5$, and 6 at the time of disk dispesal (which corresponds to the start of our simulations of the main paper). Motivated by this result, we examined single-population systems with $N_0=10$, $M_0 = 1\,M_\oplus$, and $a_\mathrm{in,0}=0.05$ au for different $k_\mathrm{MMR}$ values and compared the outcomes with the baseline simulation using $k_\mathrm{orb}=10$.

Figure~\ref{fig: Rp_Ecc_psfecc0} shows the cumulative distributions of orbital period, eccentricity, and core radius in simulations where planets are assumed to be initially in mean motion resonance and simulations where $k_\mathrm{orb}=10$ (nominal case). As expected, larger $k_\mathrm{MMR}$ values correspond to smaller initial orbital separations, resulting in more compact systems (see dashed lines in Figure~\ref{fig: Rp_Ecc_psfecc0}(a)).

Regarding eccentricity (Figure~\ref{fig: Rp_Ecc_psfecc0}(b)), we find that the final eccentricity distribution depends only weakly on the initial orbital separation. The difference in mean eccentricity relative to the $k_\mathrm{orb}=10$ case is less than $10\%$. In contrast, other parameters -- such as the number of protoplanets $N_0$ and their mass $M_0$ -- have a much stronger impact on the eccentricity distribution. Therefore, we conclude that the choice of initial orbital spacing plays only a minor role in shaping the final eccentricities of the system.

Finally, Fig.~\ref{fig: Rp_Ecc_psfecc0}(c) shows that smaller orbital separations lead to more frequent collisions, which in turn produces planets with larger radii.


\section{Photoevapolation model}\label{sec: Photoevapolation}

\begin{figure}
    \centering
    \includegraphics[width=0.6\linewidth]{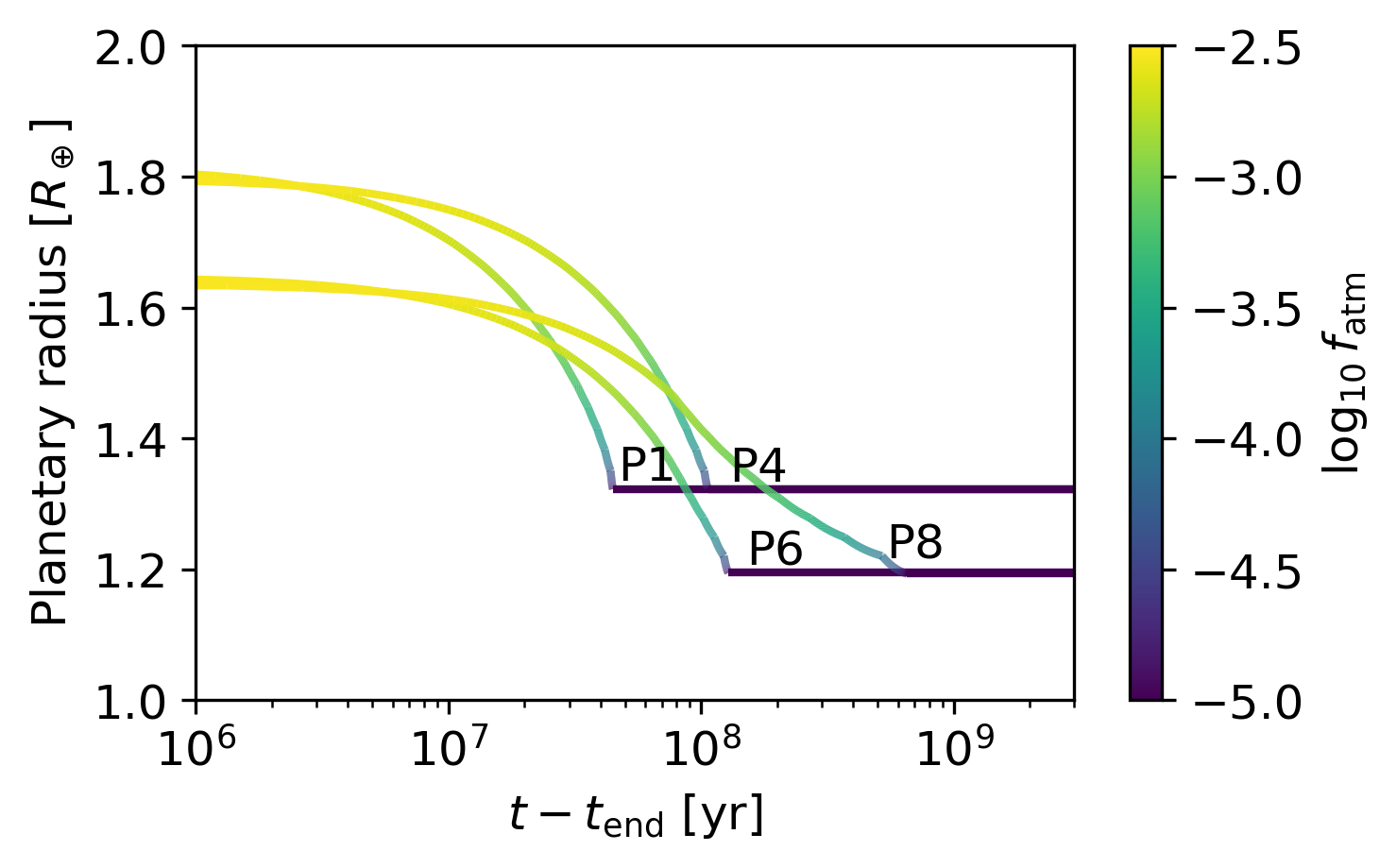}
    \caption{
    Time evolution of the protoplanet's radius. We show the same simulation shown in Fig.~\ref{fig: orb_single}. The color corresponds to the mass fraction of the planetary atmosphere. The atmospheric mass loss calculation starts from the end of the N-body simulation at $t=t_\mathrm{end}$. The IDs shown in the figure is same as those in Fig.~\ref{fig: orb_single}.
    }
    \label{fig: radius_single}
\end{figure}

We adopt the photoevaporation model developed by \citet{Owen+2017}. Figure~\ref{fig: radius_single} shows the radius evolution of planets formed in the simulation presented in Fig.~\ref{fig: orb_single}. Owing to their close-in orbits, all planets lose their primordial atmospheres through photoevaporation. As a result, every planet in this simulation becomes a super-Earth.

\section{Effects of impact induced mass loss}\label{app: impact_induced_mass_loss}

\begin{figure}
    \centering
    \includegraphics[width=0.6\linewidth]{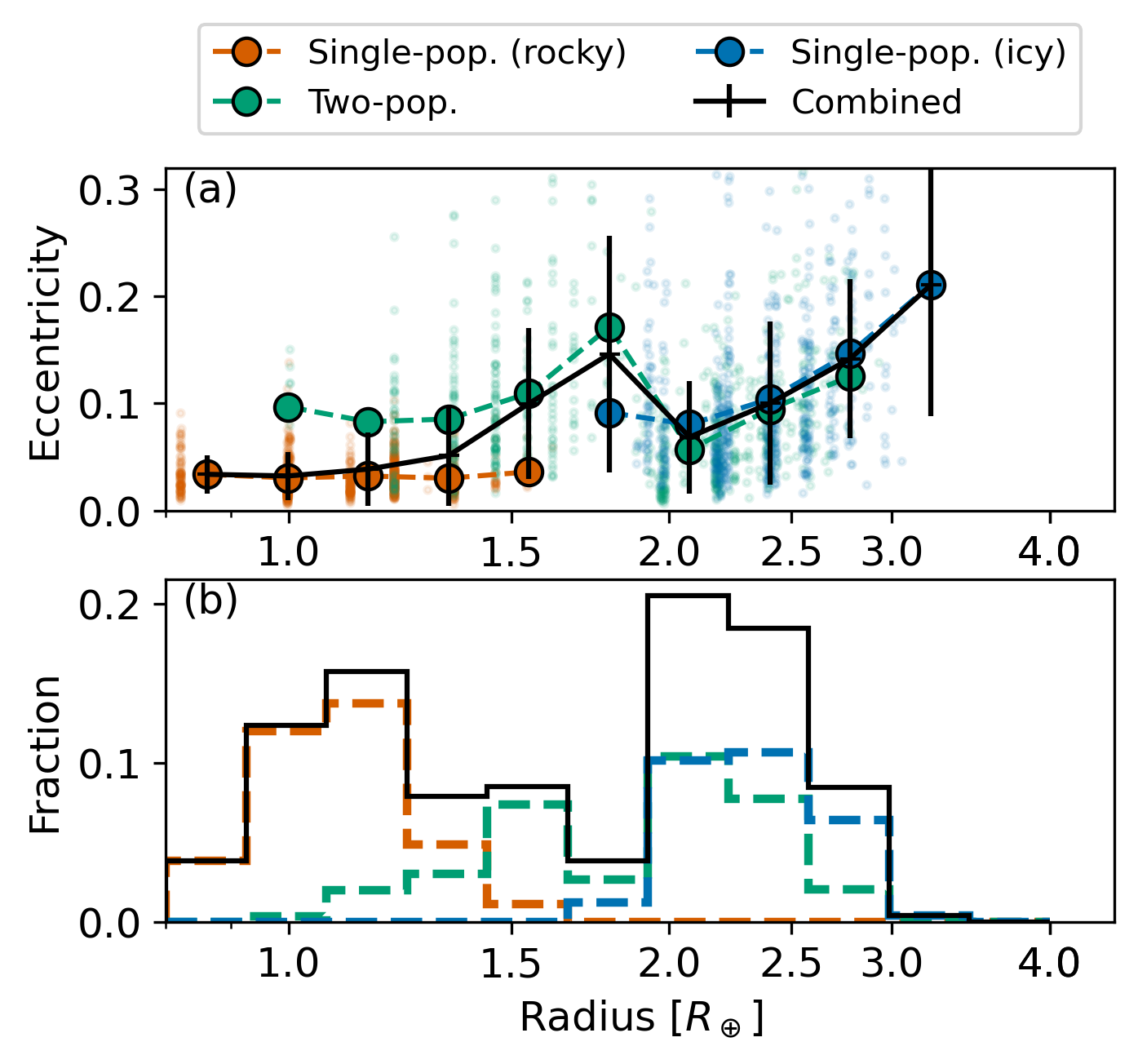}
    \caption{
    Same as Fig.~\ref{fig: Rp_ecc_mix_all}, but with the effect of impact induced mass loss.
    }
    \label{fig: Rp_Ecc_mix_all_wImp}
\end{figure}

Giant impacts among massive protoplanets can heat planetary cores and inflate their atmospheres through enhanced core luminosity, potentially triggering Parker wind–driven escape \citep{Biersteker+2019}. \citet{Biersteker+2019} showed that for impacts involving a core-mass fraction of $\gtrsim 0.1$, the resulting atmospheric loss can be complete. In our simulations, we adopt a maximally efficient prescription: any planet experiencing such an impact is assumed to lose its entire atmosphere.

As shown in Fig.~\ref{fig: Rp_Ecc_mix_all_wImp}, incorporating impact-induced atmospheric loss produces only minor changes to the overall $\langle e \rangle$–$R$ trend (compare with Figure \ref{fig: Rp_ecc_mix_all}). Rocky planets already lose nearly all of their atmospheres through photoevaporation alone, whereas icy planets remain core-dominated with radii above $2\ R_\oplus$ even if completely stripped. Consequently, our main conclusions are insensitive to the particular treatment of atmospheric mass loss.

\section{Scaled eccentricity and orbital separation}\label{app: scaling_laws}

We calculate the eccentricity and orbital separation scaled with the Safrovnov number suggested by \citet{Kokubo+2025}. They define the system's mean semi-major axis, orbital separation, and eccentricity as
\begin{align}
    a_M &= \frac{\sum_j^N M_j a_j}{\sum_j^N M_j}, \\
    \tilde{b}_\mathrm{K} &= \frac{1}{N-1} \sum_{j}^{N-1} \frac{a_{j+1}-a_j}{r_\mathrm{K,j}}, \\
    \tilde{e}_\mathrm{K} &= \frac{1}{N-1} \sum_{j}^{N-1} \frac{a_j e_j +a_{j+1} e_\mathrm{j+1}}{2 r_\mathrm{K,j}},
\end{align}
with
\begin{align}
    r_\mathrm{K,j} = \sqrt{\Theta} \frac{a_{j}+a_{j+1}}{2}.
\end{align}

\begin{figure}
    \centering
    \includegraphics[width=0.6\linewidth]{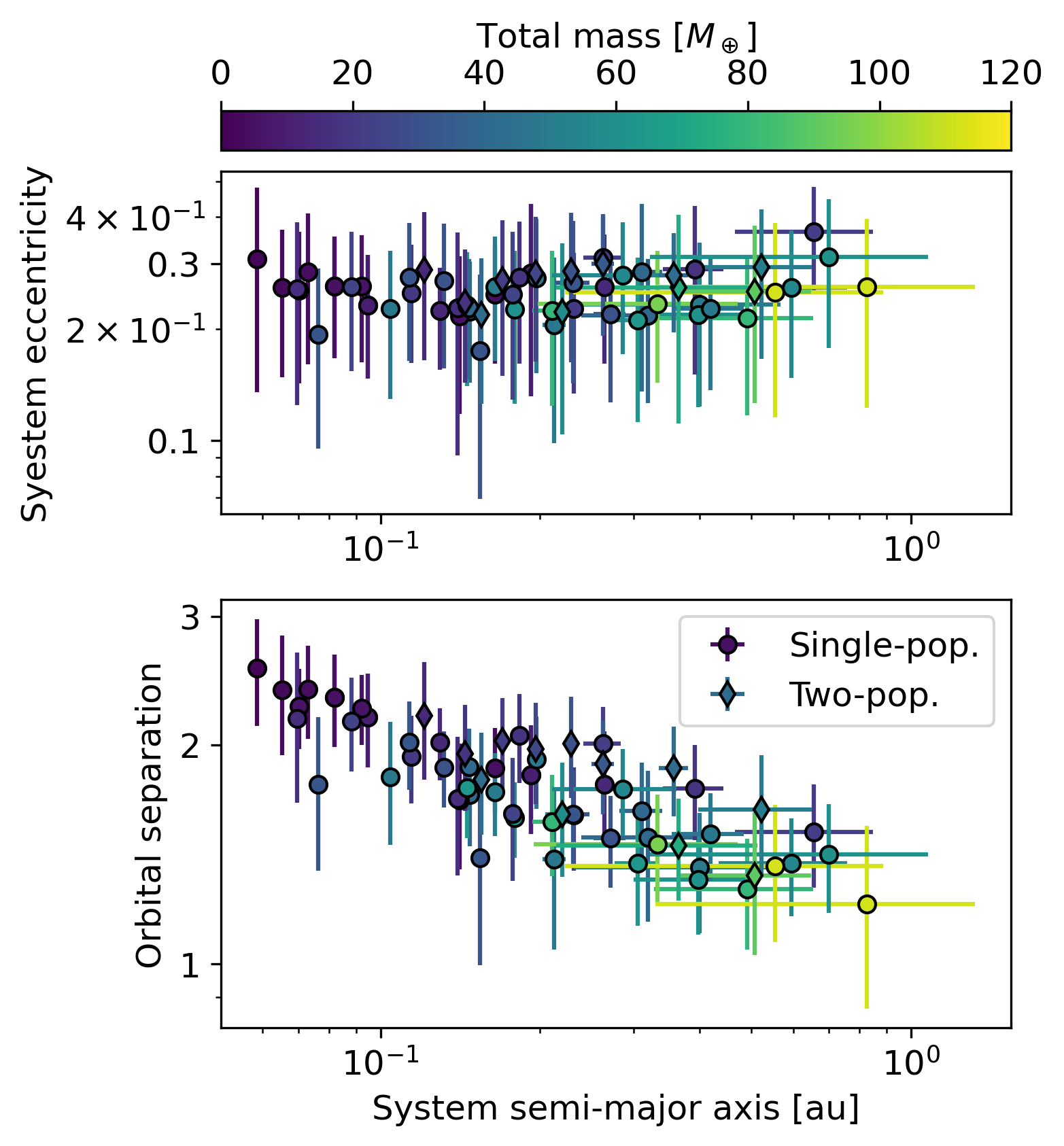}
    \caption{
    Scaled eccentricity and orbital separation obtained in our simulations.
    }
    \label{fig: system_axi_ecc}
\end{figure}

Figure~\ref{fig: system_axi_ecc} shows the mean eccentricity and orbital separation as a function of the mean semi-major axis. Single-population simulations are shown with circles, and two-population simulations with diamonds. The scaled eccentricity is nearly independent of the total mass and the mean semi-major axis. As shown in \citet{Kokubo+2025}, the scaled orbital separation decreases weakly with increasing system mean semi-major axis. Importantly, these scaling relations hold for the two-population systems as well.



\section{Fitting to system Safronov number}\label{app: fitting_Ecc_Theta}

\begin{figure}
    \centering
    \includegraphics[width=0.6\linewidth]{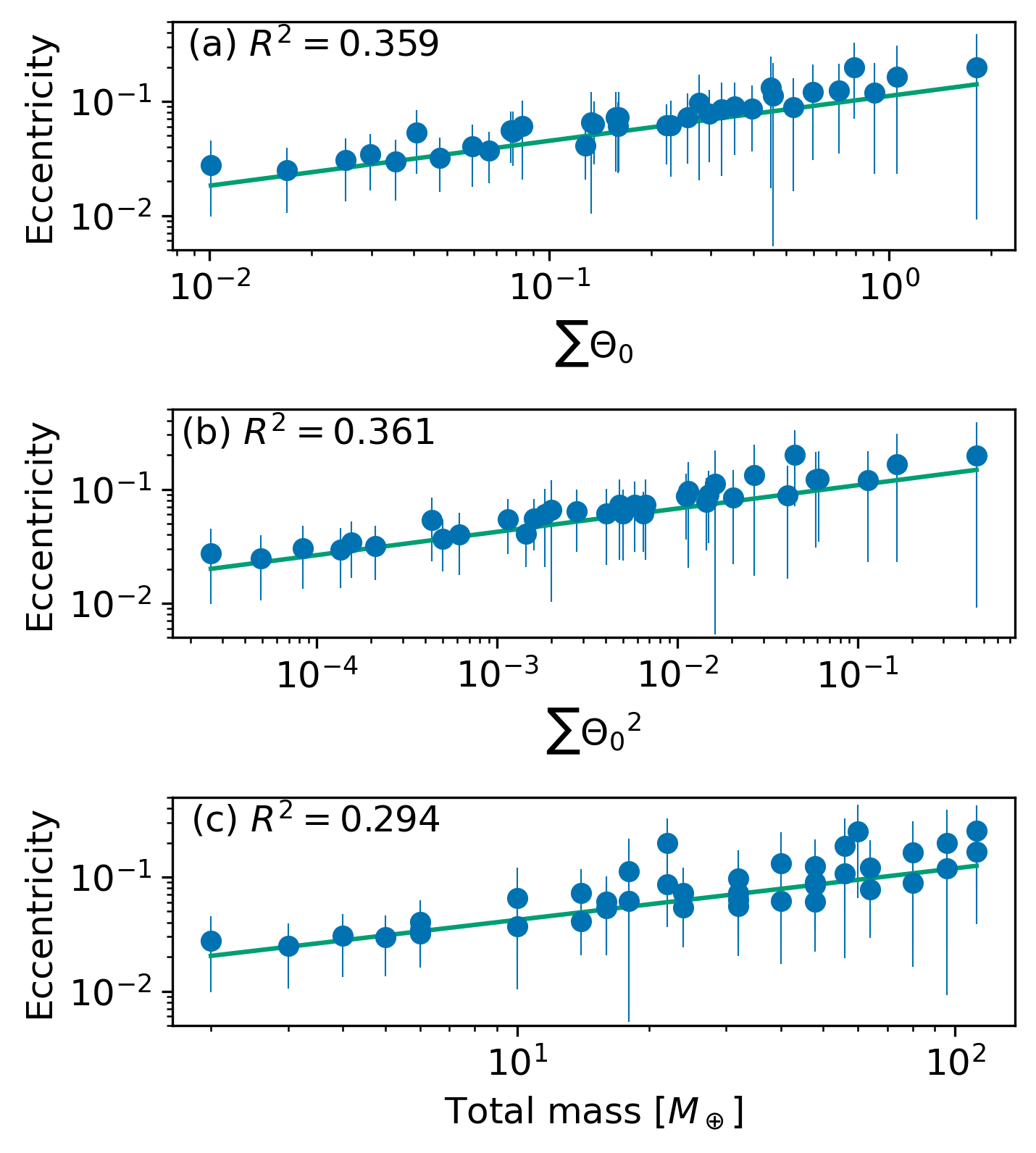}
    \caption{
    Mean eccentricity as a function of (a) $\sum \Theta_{i, 0}$, (b) $\sum {\Theta_{i, 0}}^2$, and (c) total mass of protoplanets $N_0 M_0$. We display the $R^2$ values obtained from the power-law fitting in the top left corner of each panel. Note that we show the mean eccentricity for each simulations in this figure, but we use eccentricities of each planet for the power-law fitting.
    }
    \label{fig: Ecc_Theta_fit}
\end{figure}


To further quantify the Safronov-number interpretation introduced above, we perform power-law fits between the mean eccentricity and several candidate system-level predictors. The goal of this step is not to derive a universal predictive model, but rather to test whether the system Safronov number captures the dominant physical scaling of eccentricity excitation during orbital instability. This fitting analysis should therefore be viewed as a quantitative refinement of the Safronov framework, rather than as a separate modeling effort.

We applied power-law fits to the mean planetary eccentricity using three different predictors: 
(i) the sum of Safronov numbers $\sum \Theta_{i,0}$, 
(ii) the sum of squared Safronov numbers $\sum {\Theta_{i,0}}^{2}$, and 
(iii) the total mass of the system $N_0M_0$. 
Figure~\ref{fig: Ecc_Theta_fit} shows the resulting fits, with $R^{2}$ values of 0.414, 0.410, and 0.264, respectively. 
Thus, $\sum \Theta_{i,0}$ and $\sum {\Theta_{i,0}}^{2}$ correlate noticeably better with the final eccentricities than does the total system mass.

The physical motivation for $\sum {\Theta_{i,0}}^{2}$ comes from the scaling of scattering-to-collision probability, which is proportional to ${\Theta^{2}}$ when $\Theta/e^{2} < 1$ (see Eq.~\ref{eq: lambda}). 
During orbital instability, some planets evolve from the regime $\Theta/e^{2}<1$ into $\Theta/e^{2}>1$, so both regimes occur in our simulations. 
This explains why $\sum {\Theta_{i,0}}^{2}$ performs almost as well as $\sum \Theta_{i,0}$.

In this paper, we define $\Theta_{\mathrm{sys}} = \sum \Theta_i$ as the ``system Safronov number'' because it is physically intuitive and statistically robust; however, the quantity $\sum {\Theta_i}^{2}$ may provide an even better predictor in specific dynamical regimes. Exploring this distinction further is an interesting topic for future work.

For the system Safronov number $\Theta_{\mathrm{sys},0}$, the best-fit relation is:
\begin{align}
    \log_{10} e = 0.39 \, \log_{10} \Theta_{\mathrm{sys},0} - 0.95.
\end{align}





\end{document}